\documentclass[11pt,twoside,a4paper,cmspaper,final,collab]{cms-tdr}
\def\svnVersion{b393c83}\def\svnDate{2026/10/08}\def\cmsCernNoTag{CERN-EP-2026-137}\def\cmsCernDate{\today}\def\cmsMessage{Published in Physical Review D as \href{http://dx.doi.org/10.1103/ktbp-csd3}{\doi{10.1103/ktbp-csd3}.}}
\begin{document}\cmsNoteHeader{B2G-24-013}

\newcommand{\cPT}{\PQT\xspace}
\providecommand{\PQBpr}{\HepParticle{\PB}{}{\prime}\xspace}
\providecommand{\PQTpr}{\HepParticle{\PQT}{}{\prime}\xspace}
\newcommand{\tW}{\ensuremath{\PQt\PW}\xspace}
\newcommand{\bZ}{\ensuremath{\PQb\PZ}\xspace}
\newcommand{\bH}{\ensuremath{\PQb\PH}\xspace}
\newcommand{\WW}{\ensuremath{\PW\PW}\xspace}
\newcommand{\WZ}{\ensuremath{\PW\PZ}\xspace}
\newcommand{\ZZ}{\ensuremath{\PZ\PZ}\xspace}
\newcommand{\ttH}{\ensuremath{\ttbar\PH}\xspace}
\newcommand{\ttW}{\ensuremath{\ttbar\PW}\xspace}
\newcommand{\ttZ}{\ensuremath{\ttbar\PZ}\xspace}
\newcommand{\ttX}{\ensuremath{\ttbar\PX}\xspace}
\newcommand{\ttV}{\ensuremath{\ttbar\PV}\xspace}
\newcommand{\SD}{\textrm{s.d.}}
\newcommand{\ST}{\ensuremath{S_\mathrm{T}}\xspace}
\newcommand{\mtw}{\ensuremath{m_{\tW}}\xspace}
\newcommand{\mB}{\ensuremath{m_{\PQBpr}}\xspace}
\providecommand{\ParticleNet}{\textsc{ParticleNet}\xspace}
\providecommand{\COMBINE}{\textsc{Combine}\xspace}

\ifthenelse{\boolean{cms@external}}{\providecommand{\cmsLeft}{upper\xspace}}{\providecommand{\cmsLeft}{left\xspace}}
\ifthenelse{\boolean{cms@external}}{\providecommand{\cmsRight}{lower\xspace}}{\providecommand{\cmsRight}{right\xspace}}

\cmsNoteHeader{B2G-24-013}
\title{Search for production of a single vector-like \texorpdfstring{\PQBpr}{B'} quark decaying to a top quark and a \texorpdfstring{\PW}{W} boson in the single-lepton final state in proton-proton collisions at \texorpdfstring{$\sqrt{s}=13\TeV$}{sqrt(s) = 13 TeV}}

\author[cern]{The CMS Collaboration}

\date{\today}

\abstract{
A search is presented for the production of a single narrow-width vector-like \PQBpr quark that decays to a \PQt quark and a \PW boson. This search targets a single-lepton final state, where a muon or electron originates either from the semileptonic decay of the \PQt quark or from the leptonic decay of the \PW boson, with the other particle decaying hadronically. The (semi-)leptonic portion of the event is reconstructed from an electron or muon, missing transverse momentum, and a small-radius jet identified as a \PQb quark jet if the muon or electron originates from the \PQt quark. The hadronic portion of the event is reconstructed as a single large-radius jet. The originating particle of the large-radius jet is identified using a neural-network-based tagger, and the dominant background contributions are modeled from data using a neural autoregressive flow network. The data were collected from 2016 to 2018 by the CMS experiment at the LHC in proton-proton collisions at $\sqrt{s}=13\TeV$, corresponding to an integrated luminosity of 138\fbinv. This search is the most sensitive to date to the production of single narrow-width \PQBpr quarks. It excludes electroweak-singlet \PQBpr quarks with $\Gamma/\mB=5\%$ for masses between 0.8 and 1.23\TeV. Limits are also placed on the production cross section of single \PQBpr quarks produced in association with \PQt quarks, and on the coupling factor of the \PQBpr quark to electroweak bosons.
}

\hypersetup{
pdfauthor={CMS Collaboration},
pdftitle={Search for production of a single vector-like B' quark decaying to a top quark and a W boson in the single-lepton final state in proton-proton collisions at sqrt(s) = 13 TeV},
pdfsubject={CMS},
pdfkeywords={CMS, VLQ, B2G}}

\maketitle 

\section{Introduction}\label{sec:Introduction}

Several models of physics beyond the standard model (SM) propose the existence of additional heavy fermions.
Many of these extensions are motivated by the concern that the SM requires delicate tuning of its parameters to agree with the observed mass of the Higgs boson.
To restore ``naturalness'' to the SM, some models propose that the Higgs boson is not a fundamental scalar at all energy scales~\cite{CH1,CH2}.
Such models often introduce ``vector-like'' quarks (VLQs) as new fermions whose left- and right-handed components transform identically under the EW symmetry groups, in contrast to the chiral nature of SM quarks~\cite{VLQ1,VLQ2}.
The VLQs could have masses up to several \TeVns and would reduce the divergent effects of \PQt quark quantum corrections to the observed Higgs boson mass.

\begin{figure}[hbtp]
   \centering
    \includegraphics[width=0.48\textwidth]{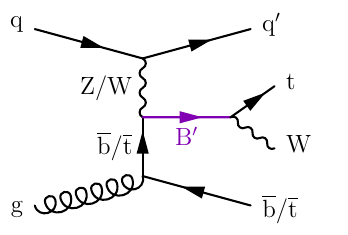}
    \caption{Tree-level Feynman diagram showing the production of a single \PQBpr quark in association with an SM \PQb quark or \PQt quark, decaying in the \tW channel.}
    \label{fig:feynman}
   \end{figure}

Typically, the VLQs are proposed to mix predominantly with third-generation SM quarks.
A vector-like bottom quark (\PQBpr) might decay to an SM \PQt or \PQb quark and a \PW, \PZ, or Higgs boson (\tW, \bZ, or \bH), thereby producing high-momentum signatures as a result of its large mass.
An example tree-level Feynman diagram for \PQBpr quark production in proton-proton ($\Pp\Pp$) collisions via an EW interaction is shown in Fig.~\ref{fig:feynman}.
The \PQBpr quark is analogous to the SM \PQb quark and has an electric charge of $-1/3\,e$, and a corresponding vector-like \PQTpr quark is proposed with an electric charge of $+2/3\,e$.
The \PQBpr quark may exist as an EW singlet, as part of a (\PQTpr, \PQBpr) doublet, or in a doublet or triplet with more exotic VLQs.
Each multiplet assignment implies different branching fractions for the decays of \PQTpr or \PQBpr quarks.
The benchmark scenarios considered in this paper are a singlet \PQBpr quark, expected to decay to \tW approximately 50\% of the time, and a \PQBpr quark in a (\PQTpr, \PQBpr) doublet that mixes only with up-type quarks. The \PQBpr quark in an up-type doublet is expected to decay exclusively to \tW~\cite{VLQ1,VLQ2} and is produced only in association with a \PQt quark.

This paper presents a search for the production of a single \PQBpr quark in association with a third-generation SM quark, where the \PQBpr decays into a \PQt quark and a \PW boson. 
The search targets single-lepton events in which either the \PQt quark decays semileptonically and the \PW boson decays hadronically, or the \PQt quark decays hadronically and the \PW boson decays leptonically. 
Consequently, the final state contains a single isolated electron or muon, missing transverse momentum, and one hadronically decaying heavy object reconstructed as a large-radius jet.
When the lepton originates from the semileptonic decay of the \PQt quark, the event also contains a small-radius \PQb-tagged jet associated with the \PQt quark decay.
This search targets the production of single \PQBpr quarks with masses in the range 0.8--2.0\TeV and decay widths for which the narrow-width approximation (NWA) is valid.
Searches targeting narrow-width VLQs probe the regime of small EW couplings and therefore provide an important complement to finite-width studies, which are primarily sensitive to larger EW couplings.
In the small-coupling regime, the interference between the signal and the irreducible SM background is negligible, allowing the signal to be modeled independently of the background~\cite{VLQ3}.
Consequently, the experimental results can be interpreted with reduced dependence on the underlying signal model.

The CMS Collaboration has conducted three searches for the production of single \PQBpr quarks at $\sqrt{s}=13\TeV$: searches for \PQBpr quarks decaying to \tW~\cite{B2G-17-018} and \bH~\cite{B2G-17-009} using a data sample corresponding to an integrated luminosity of 35.9\fbinv, and a search for excited \PQb quarks decaying to \tW using the full Run 2 data set that also provided a \PQBpr quark interpretation~\cite{B2G-20-010}.
Searches have also been conducted by the CMS Collaboration for the pair production of \PQBpr quarks via the strong interaction, excluding \PQBpr quarks decaying to \tW with masses below 1.55\TeV~\cite{exo-23-006}.
The ATLAS Collaboration has searched for \PQBpr quarks decaying to \tW at $\sqrt{s}=8\TeV$ with an integrated luminosity of 20.3\fbinv~\cite{ATLAS-Wt-8TeV}, and to \bH at $\sqrt{s}=13\TeV$ with an integrated luminosity of 139\fbinv~\cite{ATLAS-Hb-13TeV}. Of the previous searches in this channel, only Ref.~\cite{B2G-20-010} is sensitive to \PQBpr quarks with a decay width of 5\%. That search excludes \PQBpr quarks with a relative width of 5\% in the mass range 0.8--1.15\TeV (0.85--1.05\TeV) for production in association with \PQb (\PQt) quarks. 
The search presented in this paper extends these constraints, improving the upper limits on the production cross section by a factor of 2--3 for \PQb-associated production across the entire mass spectrum, and by up to a factor of 3 for \PQt-associated production for $m_{\PQBpr}>1.0\TeV$.

The remainder of this paper is organized as follows.
Section~\ref{sec:DetectorObj} introduces the CMS detector and the reconstruction of physics objects.
Section~\ref{sec:DataMC} describes the data set and simulated events.
The selection criteria and \PQBpr quark reconstruction are detailed in Sections~\ref{sec:EventSelection} and~\ref{sec:Reconstruction}, respectively.
Section~\ref{sec:BkgEstimation} describes the modeling of SM backgrounds using the ``ABCDnn'' technique, and systematic uncertainties are discussed in Section~\ref{sec:Systematics}.
Results are presented in Section~\ref{sec:Results}, and the paper is summarized in Section~\ref{sec:Summary}.
Tabulated results are provided in the HEPData record for this search~\cite{hepdata}.

\section{The CMS detector and event reconstruction}\label{sec:DetectorObj}

The CMS apparatus~\cite{CMS:2008xjf} is a multipurpose, nearly hermetic detector designed to trigger on~\cite{CMS:2020cmk,CMS:2016ngn,CMS:2024aqx} and identify electrons, muons, photons, and charged and neutral hadrons~\cite{CMS:2020uim,CMS:2018rym,CMS:2014pgm}.
Its central feature is a superconducting solenoid of 6\unit{m} internal diameter, providing a magnetic field of 3.8\unit{T}.
Within the solenoid volume are a silicon pixel and strip tracker, a lead tungstate crystal electromagnetic calorimeter (ECAL), and a brass and scintillator hadron calorimeter (HCAL), each composed of a barrel and two endcap sections.
Forward calorimeters extend the pseudorapidity ($\eta$) coverage beyond that provided by the barrel and endcap detectors.
Muons are reconstructed using gas-ionization detectors interleaved with the layers of the steel flux-return yoke outside the solenoid.
More detailed descriptions of the CMS detector, along with a definition of the coordinate system and relevant kinematic variables, can be found in Refs.~\cite{CMS:2008xjf} and~\cite{CMS:2023gfb}.

The primary $\Pp\Pp$ interaction vertex (PV) is taken to be the vertex corresponding to the hardest scattering in the event, evaluated using tracking information alone, as described in Section~9.4.1 of Ref.~\cite{CMS-TDR-15-02}.
A particle-flow (PF) algorithm~\cite{CMS:2017yfk} aims to reconstruct and identify each individual particle in an event by combining information from the various elements of the CMS detector.
Photon energies are directly measured in the ECAL.
For electrons, the momentum at the primary interaction vertex as measured by the tracker is combined with the associated ECAL cluster energy and the summed energies of bremsstrahlung photons consistent with originating from the electron track.
Muon momenta are obtained from the curvature of their reconstructed tracks.
Charged hadrons are reconstructed using their tracker-measured momenta together with matched ECAL and HCAL energy deposits, with corrections applied to account for the calorimeter response to hadronic showers.
Neutral hadrons are reconstructed from the corresponding corrected ECAL and HCAL energy measurements.

For each event, hadronic jets are clustered from PF candidates using the infrared- and collinear-safe anti-\kt algorithm~\cite{Cacciari:2008gp,Cacciari:2011ma} with a distance parameter of 0.4 (small-radius jets) or 0.8 (large-radius jets).
The jet momentum is determined as the vector sum of the momenta of its constituent particles, and it is found in simulation to agree with the true momentum to within 5--10\% across the full \pt spectrum and detector acceptance.
Additional $\Pp\Pp$ interactions within the same or nearby bunch crossings (pileup) can contribute extra tracks and calorimetric energy, increasing the apparent jet momentum.
To mitigate this effect in small-radius jets, tracks identified as originating from pileup vertices are removed, and an offset correction is applied to account for the remaining contributions~\cite{Sirunyan:2020foa}.

The pileup per particle identification algorithm (PUPPI)~\cite{Sirunyan:2020foa,Bertolini:2014bba} is used to mitigate the effect of pileup at the reconstructed particle level, making use of local shape information, event pileup properties, and tracking information. A local shape variable is defined that distinguishes between collinear and soft diffuse distributions of other particles surrounding the particle under consideration. The former is attributed to particles originating from the hard scatter and the latter to particles originating from pileup interactions. Charged particles identified as originating from pileup vertices are discarded. For each neutral particle, a local shape variable is computed from the surrounding charged particles compatible with the PV within the tracker acceptance ($\abs{\eta} < 2.5$), and using both charged and neutral particles in the region outside of the tracker coverage. The momenta of the neutral particles are then rescaled according to their probability to originate from the PV, deduced from the local shape variable, superseding the need for jet-based pileup corrections~\cite{Sirunyan:2020foa}.

Jet energy corrections are derived from simulation to ensure that the average reconstructed jet energy matches that of particle-level jets.
In situ measurements of momentum balance in dijet, \PGg{+}jet, \PZ{+}jet, and multijet events are used to determine residual differences between the jet energy scale in data and simulation, and corresponding corrections are applied~\cite{CMS:2016lmd}.
Additional quality requirements are imposed to remove jets likely dominated by instrumental effects or reconstruction failures~\cite{Sirunyan:2020foa}.
The jet energy resolution is typically 15--20\% at 30\GeV, 10\% at 100\GeV, and 5\% at 1\TeV~\cite{CMS:2016lmd}.

Events of interest are selected using a two-tiered trigger system.
The first level (L1) trigger, composed of custom hardware processors, uses information from the calorimeters and muon detectors to select events at a rate of around 100\unit{kHz} within a fixed latency of 4\mus~\cite{CMS:2020cmk}.
The second level, known as the high-level trigger (HLT), consists of a farm of processors running a version of the full event reconstruction software optimized for fast processing, and reduces the event rate to a few kHz before data storage~\cite{CMS:2016ngn,CMS:2024aqx}.

\section{Data and simulated samples}\label{sec:DataMC}

The data were collected during 2016--2018 in $\Pp\Pp$ collisions at $\sqrt{s}=13\TeV$ by the CMS experiment at the LHC, corresponding to a total integrated luminosity of 138\fbinv~\cite{CMS-LUM-17-003,CMS-PAS-LUM-17-004,CMS-PAS-LUM-18-002,CMS-PAS-LUM-20-001}.
Simulated event samples are produced separately for each data-taking year to reflect the corresponding detector conditions.

Production of a single vector-like \PQBpr quark with a narrow decay width is simulated at leading order (LO) using \MGvATNLO~\cite{Alwall:2014hca} version~2.6.5, with the model described in Refs.~\cite{theorists1,theorists2}.
The simulated \PQBpr quark masses range from 0.8 to 2.0\TeV, with a step size of 0.1\TeV for 1.2--1.8\TeV and 0.2\TeV elsewhere.
The decay width is set to 5\GeV, which is always smaller than 1\% of the mass.
The reconstructed \PQBpr quark mass has a significantly larger width than the generated width, so the reconstruction is not sensitive to small variations in the generated width as long as the width is narrow.
The results can be compared with theory predictions for VLQs with relative decay widths of 1\% and 5\%, but the interpretation is limited to widths smaller than 10\%.
The simulation includes production of a \PQt quark and a \PW boson in association with a \PQb (\PQt) quark and a light-flavor quark, denoted as $\PQb\Pq\PQBpr$ ($\PQt\Pq\PQBpr$).
The decays of the \PQt quark and \PW boson are modeled using \textsc{MadSpin}~\cite{MadSpin}.

The production cross sections of a single VLQ are computed to LO, as described in Refs.~\cite{exo-23-006,VLQ3}.
The production cross section depends on the VLQ mass, width, and couplings.
In the NWA, the cross section can be factorized into a reduced cross section $\hat{\sigma}_{\mathrm{NWA}}$ that depends only on the \PQBpr mass, and couplings of the VLQ with SM quarks and bosons~\cite{VLQ3,exo-23-006}.
The \PQb-associated (\PQt-associated) production of \PQBpr quarks is assumed to proceed primarily through a coupling to \PZ (\PW) bosons.
For a \PQBpr quark decaying to $\PQt\PW$, the production cross sections for the $\PQb\Pq\PQt\PW$ and $\PQt\Pq\PQt\PW$ signatures via a \PQBpr quark, where \Pq denotes a light-flavor quark, are given by
\begin{align}
    \sigma(\PQb\Pq\PQt\PW) &= \left(\frac{e}{2c_\text{W} s_\text{W}}\right)^2 \kappa^2 \hat{\sigma}_{\mathrm{NWA}} \mathcal{B}, \\
    \sigma(\PQt\Pq\PQt\PW) &= \left(\frac{e}{\sqrt{2} s_\text{W}}\right)^2 \kappa^2 \hat{\sigma}_{\mathrm{NWA}} \mathcal{B},
\end{align}
where $e$ is the fundamental electric charge, $c_\text{W}$ ($s_\text{W}$) is the cosine (sine) of the weak mixing angle, $\kappa$ is a ``coupling factor'' that approximates the sine of the mixing angle between the VLQ and SM quark, and $\mathcal{B}$ denotes the branching fraction for $\PQBpr\to\PQt\PW$~\cite{VLQ3,exo-23-006}.
The reduced cross section $\hat{\sigma}_{\mathrm{NWA}}$ depends only on the \PQBpr mass, while the coupling factor $\kappa$ depends on the VLQ mass, width, and EW multiplet assignment.
For the benchmark models considered in this search, the singlet scenario assumes $\kappa=\kappa_\PW=\kappa_\PZ=\kappa_\PH$, whereas the $(\PQTpr,\PQBpr)$ doublet scenario assumes $\kappa=\kappa_\PW$. 
The branching fraction $\mathcal{B}$ is fixed to 0.5 for the singlet benchmark and to unity for the (\PQTpr,\PQBpr) doublet benchmark~\cite{exo-23-006}.
The values of $\kappa$ and $\hat{\sigma}_{\mathrm{NWA}}$ used in the singlet and doublet benchmarks for \PQb-associated and \PQt-associated production are summarized in Table~\ref{tab:sigxsec}.

\begin{table}[th]
    \centering
    \topcaption{
    Values of reduced cross section $\hat{\sigma}_{\mathrm{NWA}}$ and coupling factor $\kappa$ used to compute the cross sections for the production of a \PQBpr quark that is either a singlet or a member of a (\PQTpr,\PQBpr) doublet, based on the calculations described in Refs.~\cite{exo-23-006,VLQ3}.
    The value of $\hat{\sigma}_{\mathrm{NWA}}$ depends on the \PQBpr quark mass and production mode.
    The value of $\kappa$ depends on the \PQBpr quark mass, width, and multiplet structure.
    All numerical values are rounded to two significant figures, and uncertainties due to renormalization and factorization scales are provided for $\hat{\sigma}_{\mathrm{NWA}}$.
    }
    \label{tab:sigxsec}
    \renewcommand{\arraystretch}{1.2}
    \begin{scotch}{c|cc|ccc|cc}
        \multirow{2}{*}{Mass (\TeVns)} & \multicolumn{2}{c|}{$\hat{\sigma}_{\mathrm{NWA}}$ (pb)} & \multicolumn{3}{c|}{$\kappa_\text{singlet}$} & \multicolumn{2}{c}{$\kappa_\text{doublet}$} \\
         & $\PQb\Pq\PQBpr$ & $\PQt\Pq\PQBpr$ & 1\% & 5\% & 10\% &  1\% & 10\% \\        
        \hline
        0.8  & $33^{+10}_{-7}$ & $13^{+4}_{-3}$  & 0.16  & 0.36 & 0.51 & 0.23 & 0.74 \\
        0.9  & $23^{+7}_{-5}$ & $9.8^{+2.9}_{-2.0}$  & 0.14  & 0.32 & 0.45 & 0.21 & 0.65 \\
        1.0 & $16^{+5}_{-4}$ & $7.3^{+2.2}_{-1.6}$  & 0.13  & 0.28 & 0.40 & 0.18 & 0.58 \\
        1.1 & $12^{+4}_{-3}$ & $5.5^{+1.7}_{-1.2}$  & 0.11  & 0.26 & 0.36 & 0.16 & 0.52 \\
        1.2 & $8.7^{+2.9}_{-2.0}$ & $4.2^{+1.3}_{-0.9}$  & 0.10  & 0.23 & 0.33 & 0.15 & 0.47 \\
        1.3 & $6.5^{+2.2}_{-1.5}$ & $3.2^{+1.0}_{-0.7}$  & 0.096 & 0.22 & 0.30 & 0.14 & 0.44 \\
        1.4 & $4.9^{+1.7}_{-1.1}$ & $2.5^{+0.8}_{-0.6}$  & 0.089 & 0.20 & 0.28 & 0.13 & 0.40 \\
        1.5 & $3.7^{+1.3}_{-0.9}$ & $1.9^{+0.6}_{-0.4}$  & 0.083 & 0.19 & 0.26 & 0.12 & 0.38 \\
        1.6 & $2.7^{+1.0}_{-0.7}$ & $1.5^{+0.5}_{-0.4}$  & 0.078 & 0.17 & 0.25 & 0.11 & 0.35 \\
        1.7 & $2.0^{+0.7}_{-0.5}$ & $1.2^{+0.4}_{-0.3}$  & 0.073 & 0.16 & 0.23 & 0.10 & 0.33 \\
        1.8 & $1.5^{+0.6}_{-0.4}$ & $0.94^{+0.33}_{-0.22}$ & 0.069 & 0.15 & 0.22 & 0.098 & 0.31 \\
        1.9 & $1.1^{+0.4}_{-0.3}$ & $0.74^{+0.26}_{-0.18}$ & 0.065 & 0.15 & 0.21 & 0.093 & 0.29 \\
        2.0 & $0.85^{+0.33}_{-0.22}$ & $0.59^{+0.21}_{-0.14}$ & 0.062 & 0.14 & 0.20 & 0.088 & 0.28 \\
    \end{scotch}
  \end{table} 

The \MGvATNLO generator is used at LO to simulate Drell--Yan and \PW boson production, as well as QCD multijet events.
For Drell--Yan and \PW boson production, the MLM matching scheme~\cite{Hoeche:926597} is used to remove double counting of partons between the matrix element calculations and the parton shower.
The \MGvATNLO generator is also used at next-to-leading order (NLO) in perturbative quantum chromodynamics (QCD) to simulate \ttW and \ttZ production, as well as $s$-channel single \PQt quark production.
For \ttW production, the FxFx matching scheme~\cite{Frederix:2012ps} is used to avoid double counting of jets.
The \POWHEG~v2 generator is used to simulate \ttbar~\cite{POWHEGttbar}, \ttH, and additional single \PQt quark production processes~\cite{POWHEGsingletopST,POWHEGsingletopTW} at NLO.
Production of single \PQt quarks via the $t$-channel is interfaced with \textsc{MadSpin}.
The \PYTHIA8.240 generator~\cite{Sjostrand:2014zea} is used to simulate \WW, \WZ, and \ZZ production at LO.
In figures and tables, minor background processes are grouped into ``DY+$\PV\PV$'' (containing Drell--Yan and diboson processes) and ``\ttbar{+}$X$'' (containing \ttW, \ttZ, and \ttH processes).

The NNPDF3.1~\cite{NNPDF:2017mvq} parton distribution functions (PDFs) at next-to-NLO in perturbative QCD are used to generate all background and signal samples.
Parton showering, fragmentation, and hadronization are performed using \PYTHIA with the CP5 underlying event tune~\cite{CMS:2019csb}.
The response of the CMS detector to particles is simulated using the \GEANTfour toolkit~\cite{GEANT}.
Simulated pileup interactions are included in the simulated samples.
Simulated events are weighted so that the distribution in the number of interactions per bunch crossing matches that observed in data, corresponding to an average of 23 simultaneous collisions in 2016 and 32 in 2017--2018.

\section{Event selection}\label{sec:EventSelection}

Events consistent with the decay $\PQBpr\to\PQt\PW$ are selected in a final state containing a single electron or muon, missing transverse momentum, and a hadronically decaying heavy object reconstructed as a large-radius jet.
The charged lepton and missing transverse momentum originate either from the semileptonic decay of the \PQt quark or from the leptonic decay of the \PW boson produced directly in the \PQBpr decay.
In the former case, the event also contains a small-radius jet originating from the accompanying \PQb quark from the \PQt quark decay.
Signal events must pass one of the HLT triggers for electrons, photons, or muons.
The electron triggers require either a nonisolated electron with transverse momentum $\pt>115\GeV$, or an electron with $\pt>50\GeV$ in association with a small-radius jet with $\pt>165\GeV$.
During an early segment of the 2017 data-taking period, the primary electron trigger required a high-quality isolated electron with $\pt>35\GeV$.
To maintain good efficiency for high-\pt electrons, events passing a photon trigger with an energy threshold of 175\GeV in 2016 or 200\GeV in 2017--2018 are also accepted.
The muon trigger requires a muon with $\pt>50\GeV$, and is supplemented during most data-taking periods by a trigger selecting muons with $\pt>100\GeV$ based solely on tracker information.

Leptons are selected based on their position in the detector, reconstruction quality, and isolation from nearby energy deposits.
The decay products from signal events are highly boosted because of the large mass of the \PQBpr quark, so the leptons are required to pass a tight isolation requirement on the variable $I_{\text{mini}}$~\cite{mini-Iso}, which is defined as the scalar \pt sum of charged hadron, neutral hadron, and photon PF candidates within a cone in $\eta$-$\phi$ space around the lepton, divided by the lepton \pt.
The cone radius decreases from 0.2 for leptons with $\pt\leq 50\GeV$ to 0.05 for $\pt>200\GeV$.
Selected leptons must satisfy $I_{\text{mini}}<0.1$. 
The isolation variable is corrected for pileup using an effective area method~\cite{Cacciari:2008gn}.

Electron candidates must satisfy $\abs{\eta}<2.5$, with the transition region between the barrel and endcap calorimeters ($1.442<\abs{\eta}<1.566$) excluded.
A fraction of the hadronic endcap calorimeter was not functional during a portion of the 2018 data-taking period.
Thus, electrons, small-radius jets, and large-radius jets in the inoperable detector region corresponding to $-2.5<\eta<-1.479$ and $-1.55<\phi<-0.9$ are excluded for the affected portion of the 2018 data and corresponding simulated samples.
For electrons in the barrel (endcap) region, the longitudinal impact parameter relative to the PV must be less than 0.1 (0.2)\unit{cm}, and the transverse impact parameter must be less than 0.05 (0.1)\unit{cm}.
Electron candidates must also satisfy the ``tight'' identification criteria described in Ref.~\cite{CMS:2020uim}, except that the standard isolation requirement there is replaced by the one from the previous paragraph, because the standard requirement is too restrictive. 

Muons are required to lie within the central region of the detector with $\abs{\eta}<2.4$, and to have a longitudinal (transverse) impact parameter less than 0.5 (0.2)\unit{cm} with respect to the PV.
Muon candidates must satisfy the ``medium'' identification requirement.
The overall efficiency is calculated to be 99.5\% from simulated \PW and \PZ events~\cite{CMS:2018rym}.

Signal events must contain exactly one lepton with $\pt>55\GeV$ that meets these criteria.
Events are rejected if they contain any additional leptons satisfying the selection criteria with $25<\pt<55\GeV$.

The four-momentum of selected leptons is subtracted from overlapping small-radius jets.
Jet energy scale and resolution corrections are then applied.
Selected small-radius jets must have $\pt>30\GeV$, and must be separated from the selected lepton by $\DR=\sqrt{\smash[b]{(\Delta\eta)^2+(\Delta\phi)^2}}>0.4$ or have a relative \pt greater than 20\GeV with respect to the lepton.
Small-radius jets must also satisfy a jet identification quality requirement which has an efficiency greater than 98--99\%~\cite{CMS:2016lmd}.
Jets with $\pt<50\GeV$ are more likely to originate from pileup interactions, so they must additionally satisfy a pileup-rejection criterion with efficiencies of 99\% for $\abs{\eta}<2.5$ and 95\% for $\abs{\eta}>2.5$~\cite{Sirunyan:2020foa}.
Small-radius jets are labeled as ``forward'' if they have $\abs{\eta}>2.5$.
Jets originating from \PQb hadrons are identified using the \textsc{DeepJet} algorithm with a working point that provides 98\% efficiency for true \PQb jets and a 10\% misidentification rate for light-flavor quark and gluon jets~\cite{Bols:2020bkb,CMS-DP-2018-058}.

Because of the large mass of the \PQBpr quark, the \PQt quark and \PW boson from its decay often have large Lorentz boosts, such that their decay products are collimated.
In such cases, the hadronic decay products of the \PQt quark or \PW boson can be reconstructed as a large-radius jet.
Selected large-radius jets must have $\pt>200\GeV$ and $\abs{\eta}<2.5$, and must satisfy the same jet identification requirements applied to small-radius jets.
As \PQBpr quark decays are expected to yield a high-\pt jet well separated from the lepton in $\eta$-$\phi$ space, the selected jet is required to satisfy $\DR>\pi/2$ with respect to the lepton.
Tagging of these jets as \PQt quark or \PW boson candidates is performed using the \ParticleNet algorithm~\cite{particleNet}, with working points corresponding to a misidentification rate of 1\% (5\%) for classifying light-flavor quark jets as \PQt quark (\PW boson) jets~\cite{particleNet}.
In simulated \PQBpr quark events, the efficiency for correctly identifying \PQt (\PW) jets is 85--90\% (70--90\%) for jets with $\pt > 500 \, (200)\GeV$.
A given jet may satisfy both the \PQt and \PW tagging criteria.
The choice of whether to treat a jet as a \PQt or \PW candidate follows the procedure described in Section~\ref{sec:Reconstruction}.

The missing transverse momentum vector, \ptvecmiss, is computed as the negative vector sum of the transverse momenta of all PF candidates in the event~\cite{CMS:2019ctu}.
Its magnitude is denoted as \ptmiss.
Corrections to the reconstructed jet energy scale are propagated to \ptvecmiss.
The transverse momentum of the selected lepton (denoted $\pt^{\Pell}$), \ptmiss, and the difference in azimuthal angle between the lepton and \ptvecmiss (denoted $\Delta\phi$) are used to construct the transverse mass of the \PW boson: $\mT = \sqrt{2 \pt^{\Pell} \ptmiss (1 - \cos\Delta\phi)}$.

Signal events are required to contain at least one selected large-radius jet, $\ptmiss>60\GeV$, and $\mT<200\GeV$.
The scalar sum of the \pt of all selected small-radius jets must exceed 250\GeV.
To ensure consistency with the electron trigger, events in the electron channel must contain a small-radius jet with $\pt>185\GeV$ for events with an electron of $\pt<120\GeV$.

\section{The VLQ reconstruction and event categorization}
\label{sec:Reconstruction}

\begin{figure*}[htbp]
  \centering
  \includegraphics[width=0.49\textwidth]{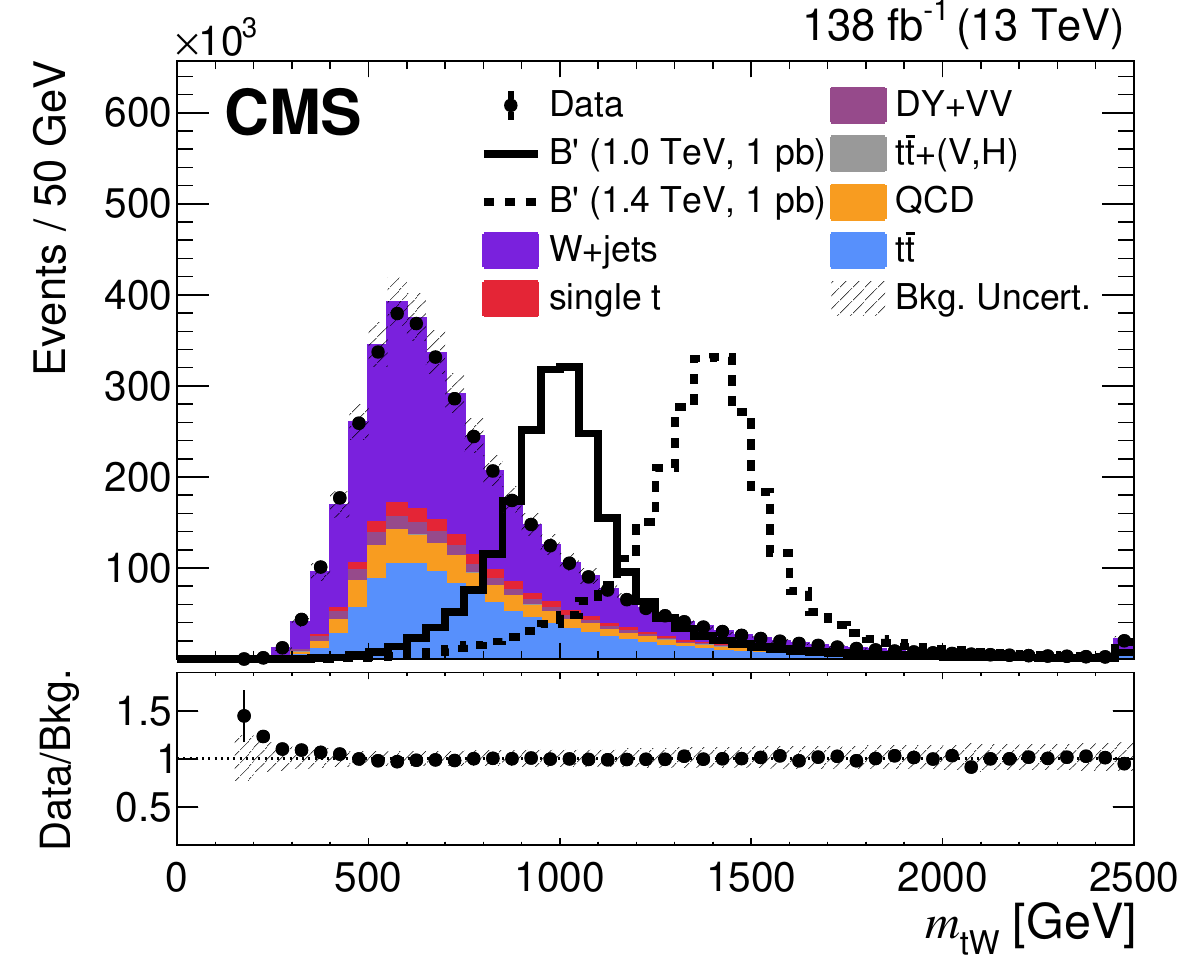}
  \includegraphics[width=0.49\textwidth]{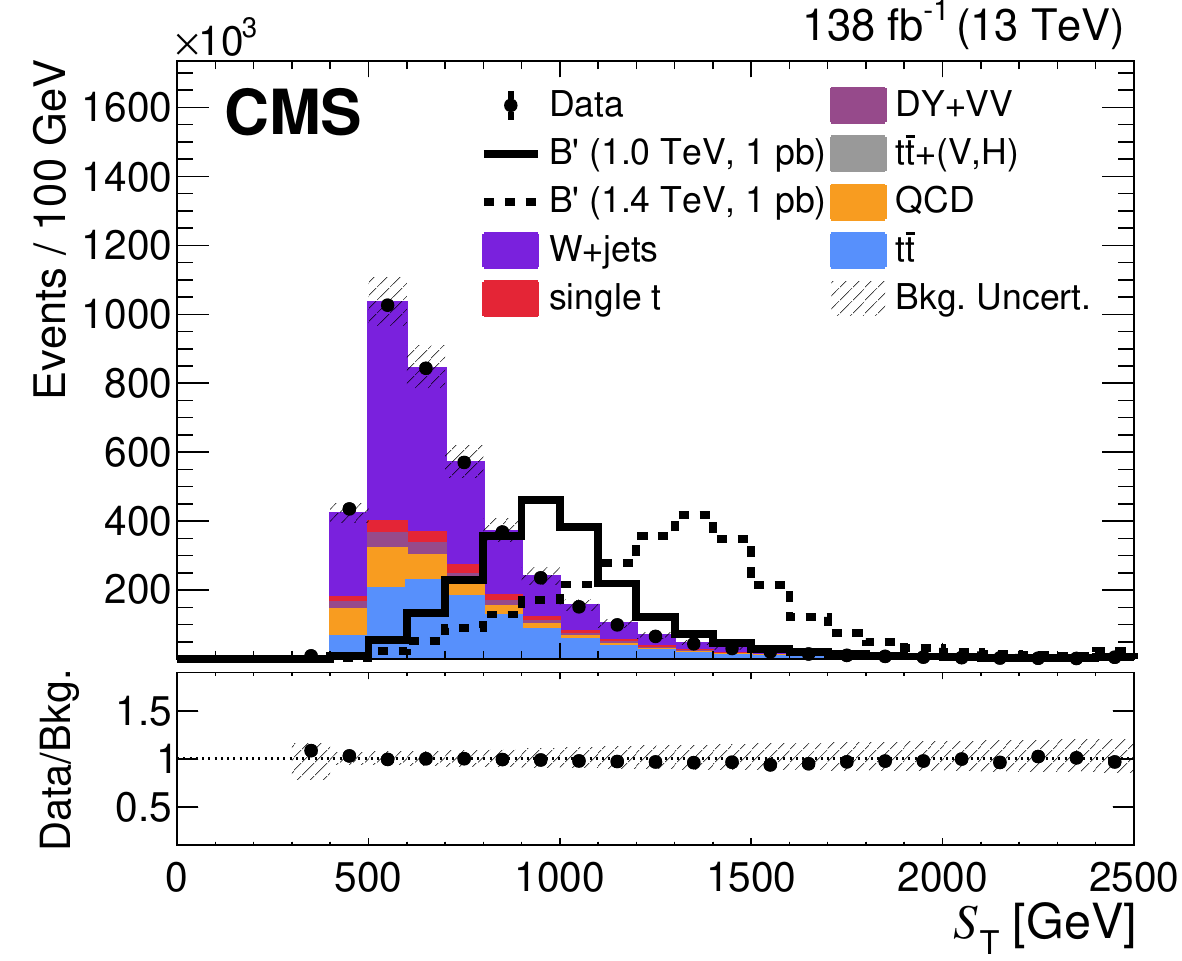}
  \includegraphics[width=0.49\textwidth]{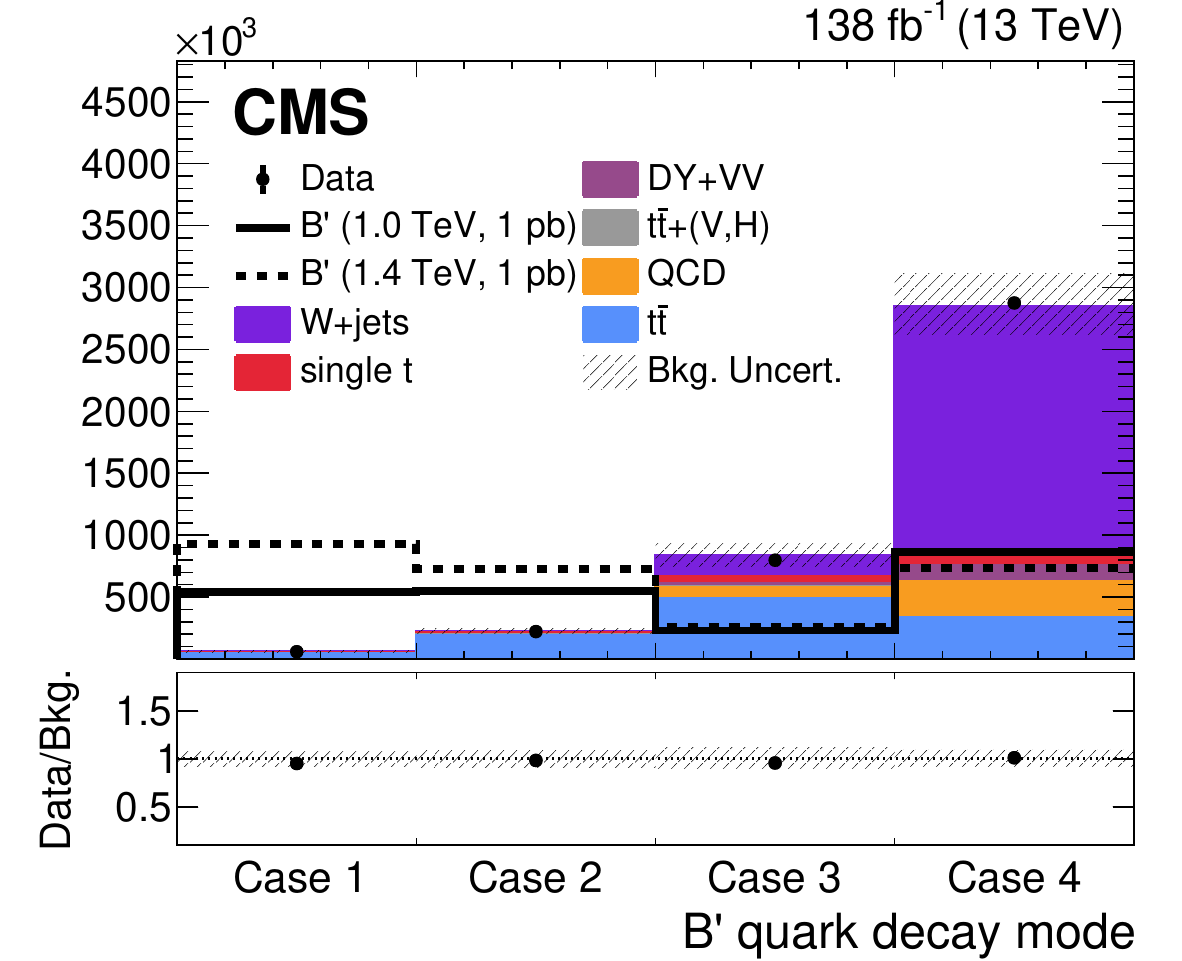}
  \caption{Distributions of the reconstructed mass of the \PQBpr quark candidate, \mtw (upper left), the scalar \pt sum (\ST) of all selected small-radius jets, the selected lepton, and \ptmiss (upper right), and the reconstruction case (lower) for all selected events.
  The observed data are shown as black markers.
  Predicted $\PQb\Pq\PQBpr$ quark signals with masses of 1.0 and 1.4\TeV are shown as solid and dashed lines, respectively, normalized to a cross section of 1\unit{pb} for visualization.
  Simulated background estimates are displayed as filled histograms.
  The lower panels show the ratio of the data to the background prediction.
  Statistical and systematic uncertainties in the background estimate are indicated by hatched bands.}
  \label{fig:mass}
\end{figure*}

The \PQBpr quark is assumed to decay to \tW, with the charged lepton originating from either $\PW \to \Pell \PGn$ decay or $\PQt \to \PQb \PW \to \PQb \Pell \PGn$ decay.
A \PW boson candidate is therefore reconstructed in all events from the charged lepton and \ptvecmiss, applying a \PW mass constraint.
The $z$ component of the neutrino momentum is taken as the real part of the solution that yields a \PW boson mass closest to the nominal value of 80.4\GeV.
If a \PQb-tagged small-radius jet is found within $\DR<\pi/2$ of the lepton, the event is categorized as a semileptonic \PQt quark event and the \PQt candidate is reconstructed from the \PW boson and the \PQb-tagged jet.
If no such \PQb-tagged jet is found, the event is categorized as a leptonic \PW boson event.
The other decay product of the \PQBpr quark, which does not produce a lepton, is assumed to correspond to the selected large-radius jet with the highest \pt.
Based on the \ParticleNet identification of this large-radius jet, events are assigned to one of four reconstruction categories:
\begin{itemize}
\item Case 1: leptonic \PW boson and a \PQt-tagged large-radius jet,
\item Case 2: semileptonic \PQt candidate and a \PW-tagged large-radius jet,
\item Case 3: semileptonic \PQt candidate and a large-radius jet that is not \PW tagged,
\item Case 4: leptonic \PW boson and a large-radius jet that is not \PQt tagged.
\end{itemize}
Cases 1 and 2 provide the highest sensitivity to a \PQBpr quark signal, while Cases 3 and 4 are dominated by background.
Figure~\ref{fig:mass} shows the distribution of the reconstructed mass (\mtw) of the \PQBpr quark candidate, the scalar \pt sum (\ST) of all selected small-radius jets, the selected lepton, and \ptmiss, and the event yields in each case.
The resolution of the \mtw distribution for signal events is approximately 10\% of the generated \PQBpr quark mass.

\begin{figure}[htbp]
  \centering
  \includegraphics[width=0.49\textwidth]{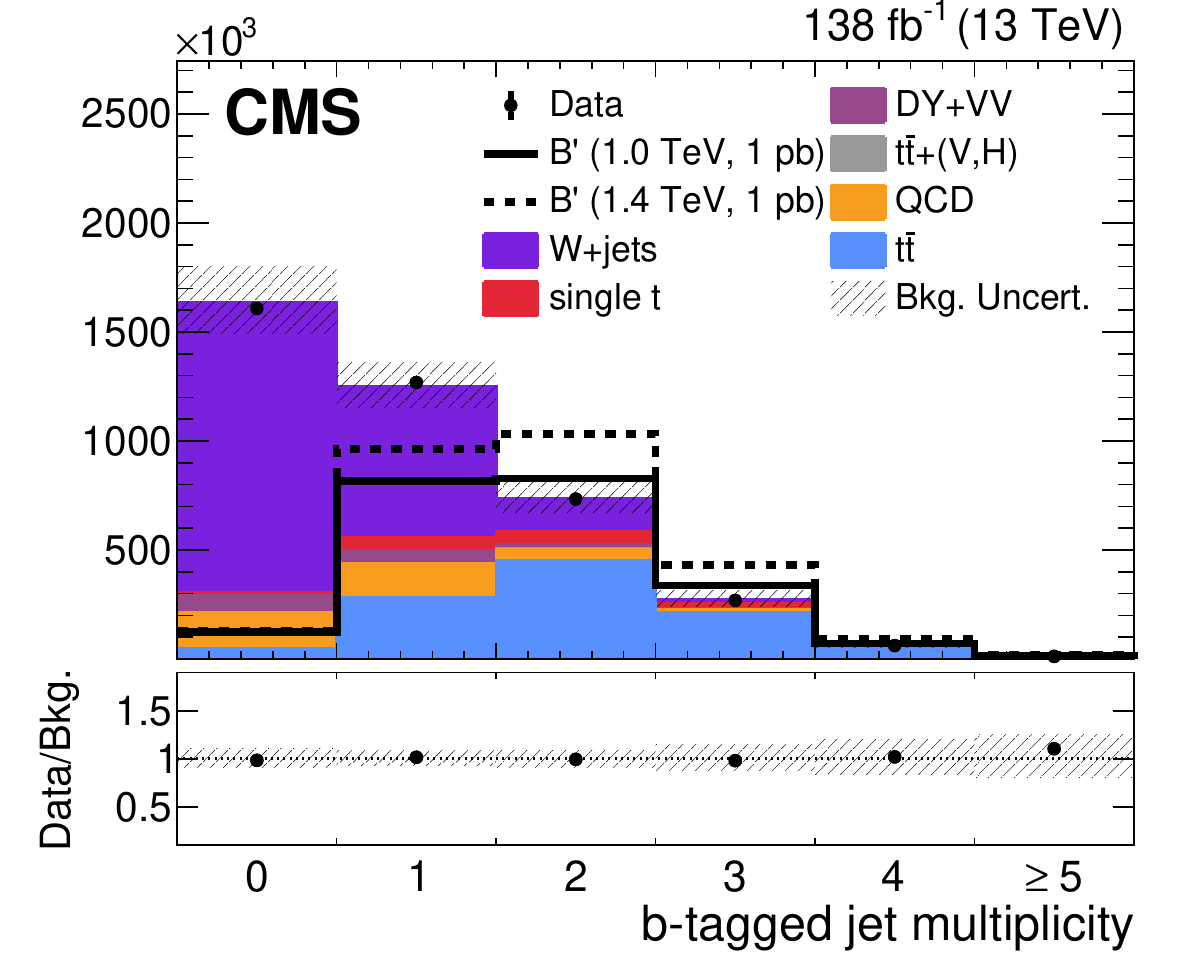}
  \includegraphics[width=0.49\textwidth]{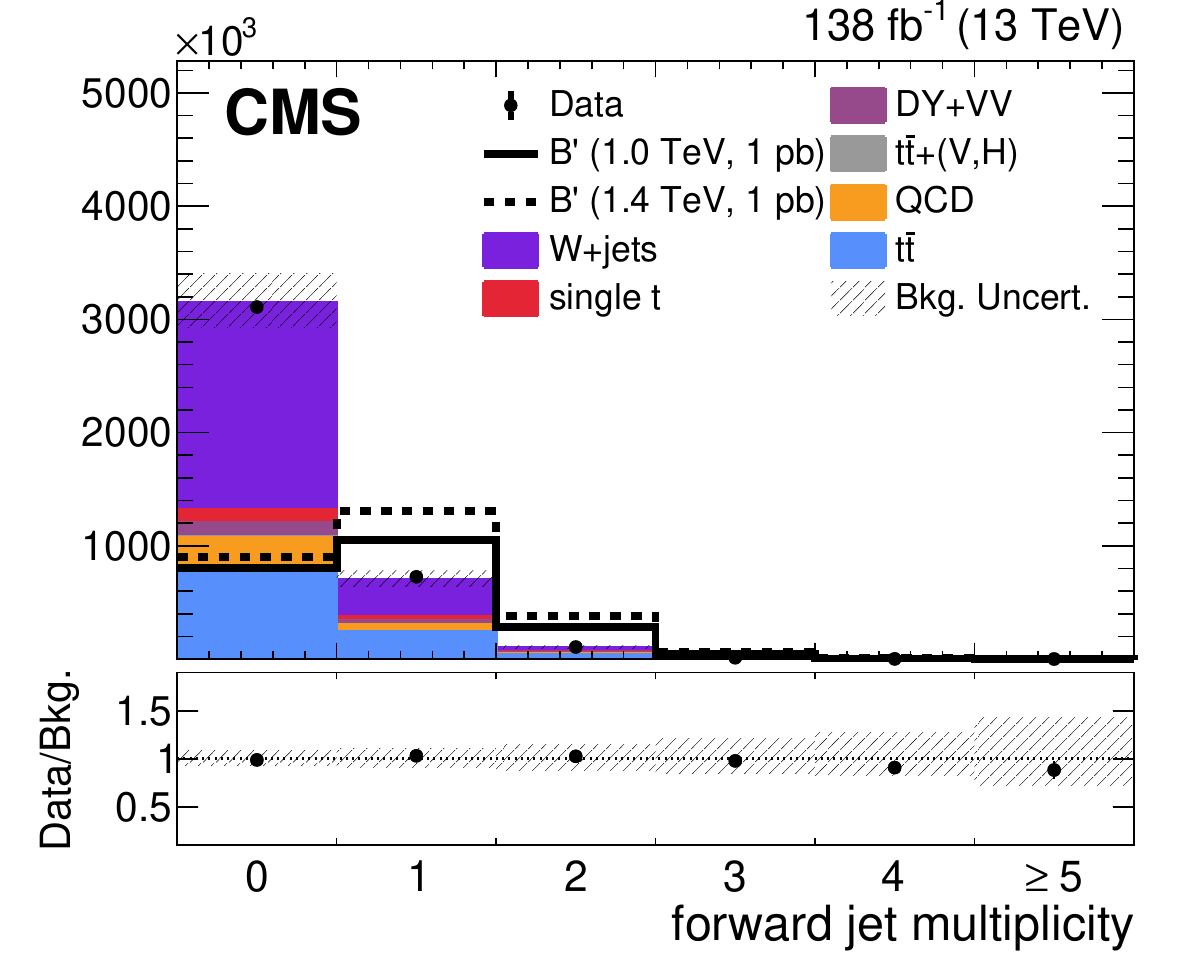}
  \caption{Distributions of \PQb-tagged jet multiplicity (\cmsLeft) and forward jet multiplicity (\cmsRight) for all selected events.
  The observed data are shown as black markers.
  Predicted $\PQb\Pq\PQBpr$ quark signals with masses of 1.0 and 1.4\TeV are shown as solid and dashed lines, respectively, normalized to a cross section of 1\unit{pb} for visualization.
  Simulated background estimates are displayed as filled histograms.
  The lower panels show the ratio of the data to the background prediction.
  Statistical and systematic uncertainties in the background estimate are indicated by hatched bands.
  }
  \label{fig:jets}
\end{figure}

\begin{figure*}[ht]
  \centering
  \includegraphics[width=0.8\textwidth]{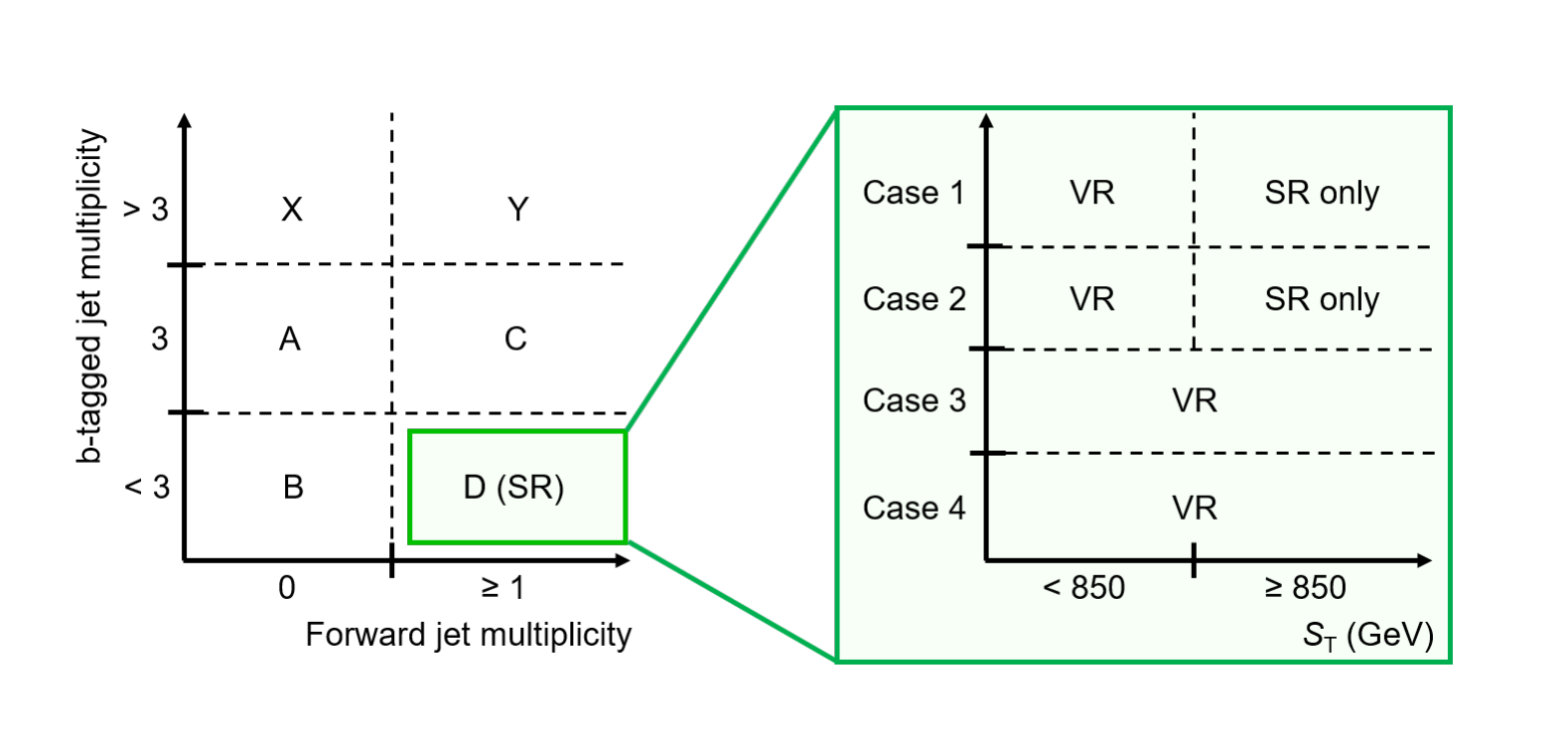}
  \caption{Diagrams illustrating the definitions and labels of the CRs and SR (left), and the subsets of the SR used for validation (right).}
  \label{fig:regions}
\end{figure*}

In \PQBpr production in association with \PQb (\PQt) and light quarks, signal events tend to have more \PQb-tagged jets than the \PW{+}jets background and more forward jets than SM backgrounds in general.
As shown in Fig.~\ref{fig:jets}, \PQBpr quark signal events tend to contain fewer \PQb-tagged jets than the \ttbar background and a larger number of forward jets than any background process.
Events are categorized into five control regions (CRs) and one signal region (SR) based on the numbers of \PQb-tagged and forward small-radius jets, as illustrated in Fig.~\ref{fig:regions} (left).
The two observables that define the regions are called ``control variables''. 
The signal is concentrated in region $D$, whereas the CRs, labeled $A$, $B$, $C$, $X$, and $Y$, are dominated by background and are used to estimate the background distribution in region $D$, as described in Section~\ref{sec:BkgEstimation}.

This categorization is optimized for \PQb-associated production.
For \PQt-associated production, region $D$ remains the most signal-enriched region, but signal events are more likely to fall into region $C$ than for \PQb-associated production.
This occurs when the \PQb quark from the associated \PQt quark lies in the central detector region. Additional central leptons or \PQb-tagged jets from the associated \PQt quark
also reduce the accuracy of the signal reconstruction, broadening the \mtw distribution in $\PQt\Pq\PQBpr$ signals.

A low signal purity portion of the SR is used to construct validation regions (VRs) to assess the modeling of background processes.
For Cases~1 and~2, the VR is defined as events in region $D$ with $\ST<850\GeV$.
For Cases~3 and~4, the VR corresponds to the full region $D$ because of the lower signal purity in these categories.
Figure~\ref{fig:regions} (right) illustrates the definition of the VRs. The final fit to data will include all events in region $D$.

\section{Background estimation}
\label{sec:BkgEstimation}

Relevant SM background processes include \ttbar, \PW{+}jets, QCD multijet, single \PQt quark, Drell--Yan, diboson, \ttV, and \ttH production.
Processes in the DY{+}$\PV\PV$ and \ttX groups, defined in Section~\ref{sec:DataMC}, comprise less than 4\% of the total background in each reconstruction case. Therefore, they are minor background sources and are modeled using simulation.
The remaining processes constitute the major background and are predicted using a combination of data-driven shape and yield estimation methods.

The shape of the \mtw distribution for the major background processes is modeled using a method called ``ABCDnn''~\cite{ABCDnn}, based on a neural network.
This method was first used in a previous CMS search~\cite{TOP-21-005}.
The ABCDnn network uses information from the five CRs, defined by two control variables, to learn transformations on one or more observables of interest and, as a result, maps their input distributions in simulation to the corresponding distributions in data.
The variables, whose distributions are modified by the transformations, are referred to as transform variables.

This technique is an extension of the traditional ``ABCD'' method, in which two control variables are used to define four regions of interest, one of which is the SR.
In the traditional method, a transfer factor is calculated from the ratio of the data yields in two neighboring CRs and applied to scale the data distribution in the remaining CR into a background prediction for the SR.
This method requires the two control variables to be independent for the background processes.
As shown in Ref.~\cite{extABCD}, predictions from the ABCD method can be improved, and the requirements on the independence of the observables weakened, by extending the method to use a larger number of regions.

This method is effective for estimating the overall background yield in the SR.
However, a bin-by-bin extension may not accurately predict the distribution of an observable when residual correlations between the control variables vary as a function of that observable.
Rather than transforming CR data into an SR background prediction using an analytic formula, the ABCDnn neural network employs neural autoregressive flows to learn a transformation that depends on the region defined by the control variables.
This transformation is applied to the simulated values of a transform variable to reproduce a distribution that matches with the data in the CRs, and to extrapolate the distribution of the transform variable into the SR. 
Potential inadequacies in the extrapolation are covered by the systematic uncertainties described in Section~\ref{sec:Systematics}.

For this search, the transform variables are \mtw and \ST.
The \mtw variable is used as the primary discriminant for signal extraction, because it is highly sensitive to the presence of signal.
Including \ST in the transformation allows the transformed \ST values to be used to define the VRs described in Section~\ref{sec:Reconstruction}.
For both transform variables (expressed in units of \GeVns), the natural logarithm is taken to compress the dynamic range of the input distributions. 

The transformation principle is expressed as:
\begin{equation}
  p_{\text{data}}(\vec{x}^\prime|\vec{c}) = \mathcal{T}(\vec{x}^\prime;\vec{x}|\vec{c}) \ast p_{\text{sim}}(\vec{x}|\vec{c}),
\end{equation}
where $p_{\text{data}}$ and $p_{\text{sim}}$ denote the distributions of the observable in data and simulation, respectively; 
$\vec{x}$ and $\vec{x}^\prime$ are vectors of (\mtw, \ST) with the values coming from the original and transformed variable values, respectively; 
and $\vec{c}$ is a vector of length 6 that identifies the region~\cite{ABCDnn}. The first through sixth elements of $\vec{c}$ correspond to regions A through Y, with the element corresponding to an event's region being set to 1, and all the other elements being set to 0.
The $\ast$ symbol denotes the convolution operation. During the training, the network learns the transformation function $\mathcal{T}(\vec{x}^\prime;\vec{x}|\vec{c})$ with $\vec{c}$ taking values corresponding to one of the CRs.

The learned transformation $\mathcal{T}(\vec{x}^\prime;\vec{x}|\vec{c})$ is then applied to the simulated major backgrounds in the SR with $\vec{c}$ taking the value of (0,0,0,1,0,0), 
which denotes region $D$. This application produces transformed values of \mtw and \ST in each simulated event, which are accumulated into a transformed background distribution.
The validity of this extrapolation is confirmed by comparing the prediction and data in the VR and is found to show bin-by-bin agreement within 20\%.

The transformation is learned using neural autoregressive flows, which combine an autoregressive conditioner network with a transformer neural network~\cite{NAF}.
The conditioner network encodes information about the transform variables and the region indices corresponding to the CRs and SR.
The transformer is implemented as a deep sigmoidal flow~\cite{NAF} that acts on the values of $\ln(\mtw/\GeVns)$ and $\ln(\ST/\GeVns)$.
Two ABCDnn models are trained separately on pairs of reconstruction cases that share the same leptonic decay channel: one for events with leptonic \PW bosons (Cases 1 and 4, referred to as the ``lepW'' model), and one for events with semileptonic \PQt quarks (Cases 2 and 3, referred to as the ``lepT'' model).
With this approach, the signal contamination across all CRs in the two training samples is no larger than 0.3\%, assuming the predicted $\PQb\Pq\PQBpr$ production cross sections at a relative decay width of $\Gamma/m_{\PQBpr} = 5\%$.

During each training epoch, one CR is selected at random, and the network is updated using samples of data and simulated distributions from that region.

For both the lepW and lepT models, the conditioner network uses the rectified linear unit activation function~\cite{relu}, and the transformer network uses a softplus activation function~\cite{softplus}.
The numbers of hidden layers and nodes per layer in each network are optimized using a random search procedure.
The loss function for training the transformer is based on the maximum mean discrepancy (MMD)~\cite{mmd}, defined as:
\begin{widetext}
\begin{equation}
    \mathrm{MMD} = \frac{1}{m(m-1)} \sum_{i} \sum_{j \neq i} k(\mathbf{x_i}, \mathbf{x_j})
    - \frac{2}{m^2} \sum_{i} \sum_{j} k(\mathbf{x_i}, \mathbf{y_j})
    + \frac{1}{m(m-1)} \sum_{i} \sum_{j \neq i} k(\mathbf{y_i}, \mathbf{y_j}),
\end{equation}
\end{widetext}
where $m$ is the number of data points, $\mathbf{x_i}$ is a vector that denotes observed (\mtw, \ST) in data, $\mathbf{y_i}$ denotes the corresponding vector predicted by ABCDnn, and $k(\mathbf{x_i}, \mathbf{x_j})$ is the radial basis function (RBF) kernel.
The RBF kernel is defined as:
\begin{equation}
    k(\mathbf{x_i}, \mathbf{x_j}) = \re^{-\|\mathbf{x_i} - \mathbf{x_j}\|^2 / 2s^2},
\end{equation}
where $\|\mathbf{x_i} - \mathbf{x_j}\|$ is the distance between $\mathbf{x_i}$ and $\mathbf{x_j}$, defined as $\mathbf{x_i}^\intercal \mathbf{x_i} - 2 \mathbf{x_i}^\intercal \mathbf{x_j} + \mathbf{x_j}^\intercal \mathbf{x_j}$, and the scaling constant $s$ controls the tolerance for differences between the two distributions.
When the data distribution and the ABCDnn-generated distribution are identical, the MMD evaluates to zero.
Larger values of $s$ have a higher tolerance for differences between the two distributions and capture global distribution shifts, while smaller values capture fine-grained differences.
Taking the average of the MMD calculations with different $s$ values allows the loss function to capture differences across multiple scales and improves the robustness of the network.
In this search, the loss function is defined as the average of three MMD calculations, each using a different $s$ value.
The number of $s$ values and their optimal values are determined using a random search.

\begin{figure*}[htbp]
  \centering
  \includegraphics[width=0.9\textwidth]{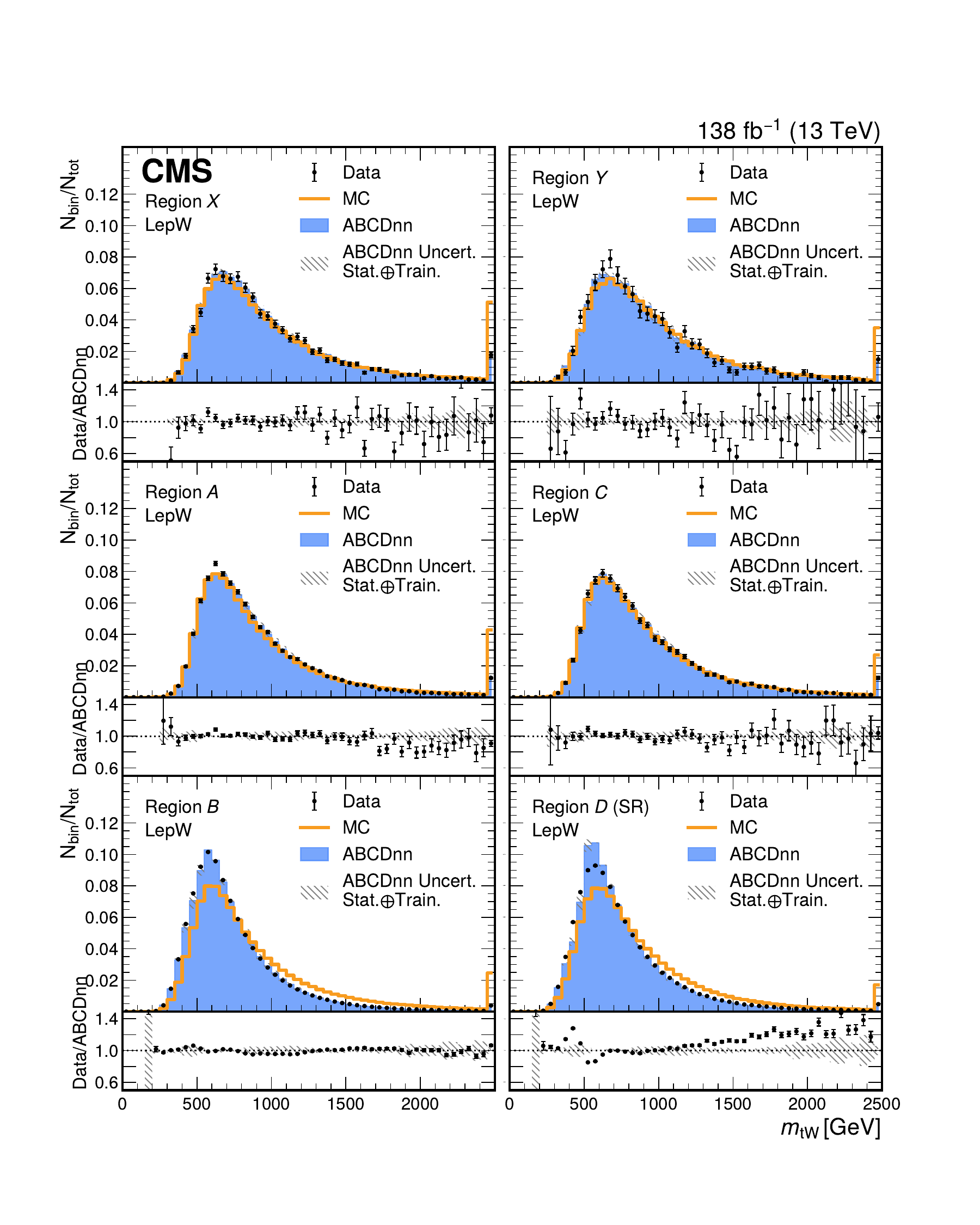}
  \caption{
  Distributions of \mtw in the SR and CRs for the lepW ABCDnn model, normalized to unity.
  Observed data are shown as black markers with statistical uncertainties.
  Major background distributions from simulation are shown as orange solid lines with statistical uncertainties and are labeled ``MC''.
  The ABCDnn predictions are shown as blue filled histograms.
  The lower panels show the ratio of the data to the ABCDnn prediction.
  Statistical and training uncertainties in the ABCDnn predictions are indicated by hatched bands, though they are too small to be visible in the upper panels.
  }
  \label{fig:training_lepW}
\end{figure*}

\begin{figure*}[htbp]
  \centering
  \includegraphics[width=0.9\textwidth]{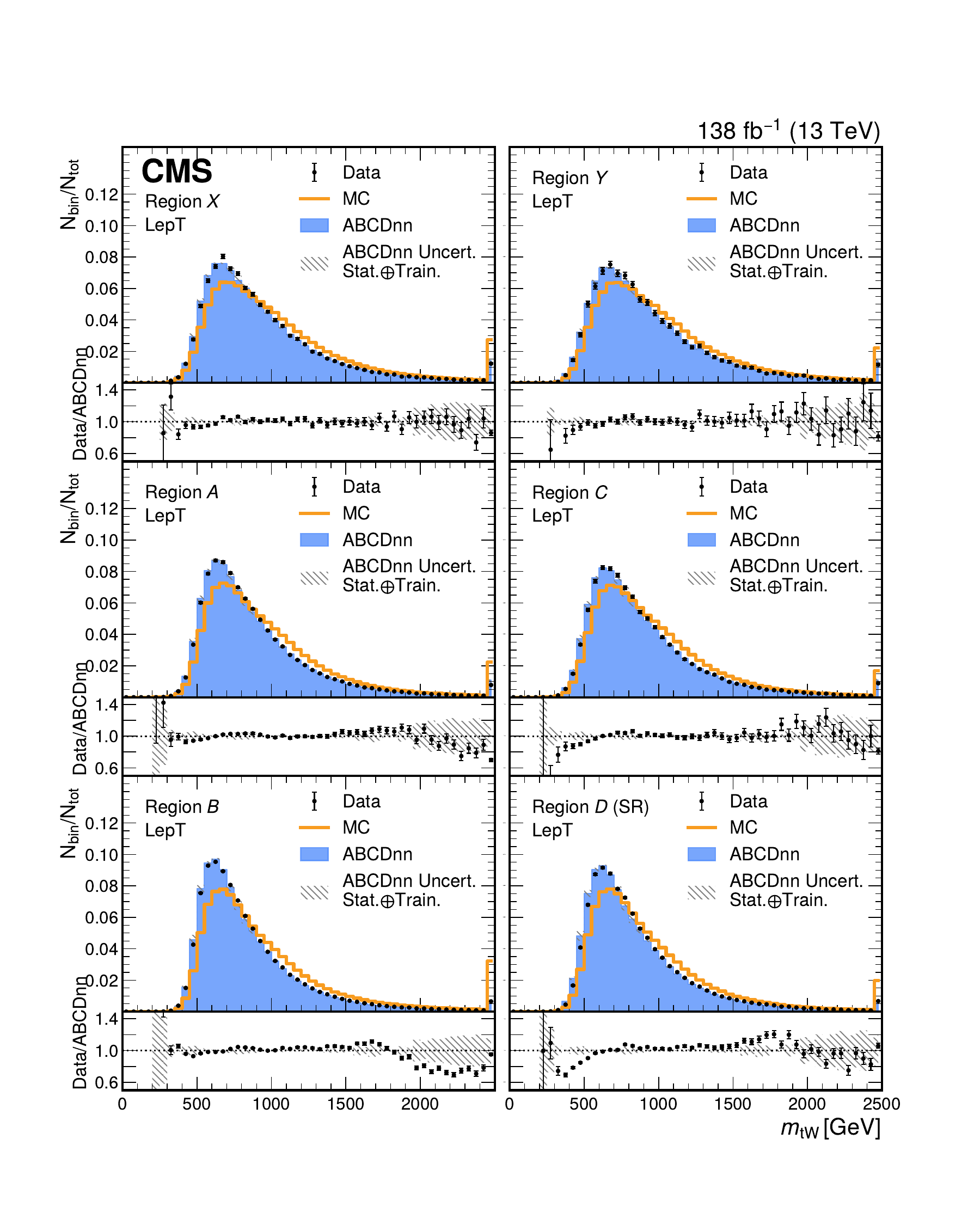}
  \caption{
  Distributions of \mtw in the SR and CRs for the lepT ABCDnn model, normalized to unity.
  Observed data are shown as black markers with statistical uncertainties.
  Major background distributions from simulation are shown as orange solid lines with statistical uncertainties and are labeled ``MC''.
  The ABCDnn predictions are shown as blue filled histograms.
  The lower panels show the ratio of the data to the ABCDnn prediction.
  Statistical and training uncertainties in the ABCDnn predictions are indicated by hatched bands, though they are too small to be visible in the upper panels.
  These figures show that the network is able to transform the unweighted distribution of the simulated major background events (labeled ``MC'') to model the data in each region.
  }
  \label{fig:training_lepT}
\end{figure*}

Model parameters are optimized by evaluating discrepancies between the data and the predicted \mtw distributions in the CRs.
Comparisons are performed using the full CR sample, as well as subsets of SR events with $\ST<850\GeV$.
The architectures and the RBF kernel sizes of the lepW and lepT models 
are summarized in Table~\ref{table:architecture}.
Figures~\ref{fig:training_lepW} and \ref{fig:training_lepT} show the training performance of the lepW and lepT ABCDnn models, respectively, for \mtw.
These figures show that the network is able to transform the unweighted distribution of the simulated major background events (labeled ``MC'') to model the data in each region.
Agreement between the data and the predictions, as functions of \mtw in the CRs, is within 20\% and differences are treated as systematic uncertainties, as described in Section~\ref{sec:Systematics}.
The training performance for the \ST transform variable is evaluated in two bins, $\ST<850\GeV$ and $\ST>850\GeV$, to validate the VR defined using this threshold.
For \ST, the discrepancies between data and prediction in the five CRs are approximately 5\% in regions far from the SR (e.g., region $X$), and approximately 1\% in regions close to the SR (e.g., region $B$).

\begin{table}[thp]
  \centering
  \topcaption{
  The ABCDnn architectures and RBF kernel sizes for the lepW and lepT models.
  The lepW model is trained on events in reconstruction Cases~1 and~4, and the lepT model is trained on events in reconstruction Cases~2 and~3.
  }
  \label{table:architecture}
  \begin{scotch}{lcc}
      Parameter & lepW Model & lepT Model \\
      \hline
      Hidden layers (conditioner net) & 14 & 11 \\
      Nodes per layer (conditioner net) & 5 & 10 \\
      Hidden layers (transformer net) & 4 & 6 \\
      Nodes per layer (transformer net) & 13 & 8 \\
      RBF kernel $s_1$ & 0.02 & 0.43 \\
      RBF kernel $s_2$ & 0.39 & 0.60 \\
      RBF kernel $s_3$ & 0.68 & 0.70 \\ 
  \end{scotch}
\end{table}

For each of the Cases 1--4, the normalization of the ABCDnn prediction in the SR is defined as
\begin{equation}\label{eq:norm}
  N^{\text{bkg}}_{\text{SR}} = N^{\text{data}-\text{minor}}_{B} \frac{N^{\text{sim}}_{\text{SR}}}{N^{\text{sim}}_{B}},
\end{equation}
where $N^{\text{data}-\text{minor}}_{B}$ is the number of events in region $B$ observed in data after subtraction of the minor background contributions estimated from simulation, and $N^{\text{sim}}_{B}$ is the number of events in region $B$ in the major background simulation.
Region $B$ is used for this normalization because it is adjacent to the SR and is well populated.
The normalization factor is determined separately for each reconstruction case.
For VRs defined with $\ST<850\GeV$, the same \ST requirement is applied to data and simulation in region $B$ when applying Eq.~\ref{eq:norm}.

\begin{figure*}[htbp]
  \centering
  \includegraphics[width=0.49\textwidth]{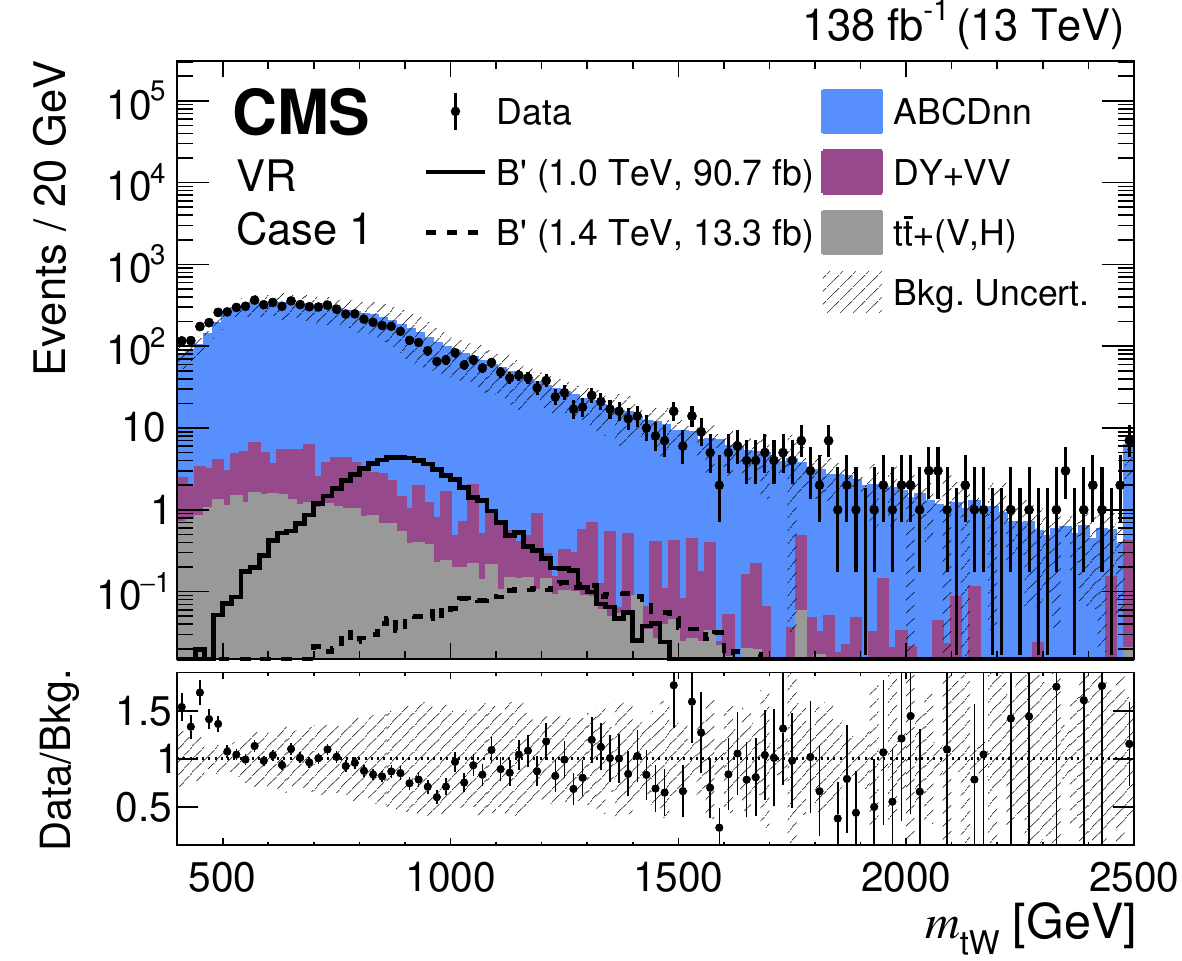}
  \includegraphics[width=0.49\textwidth]{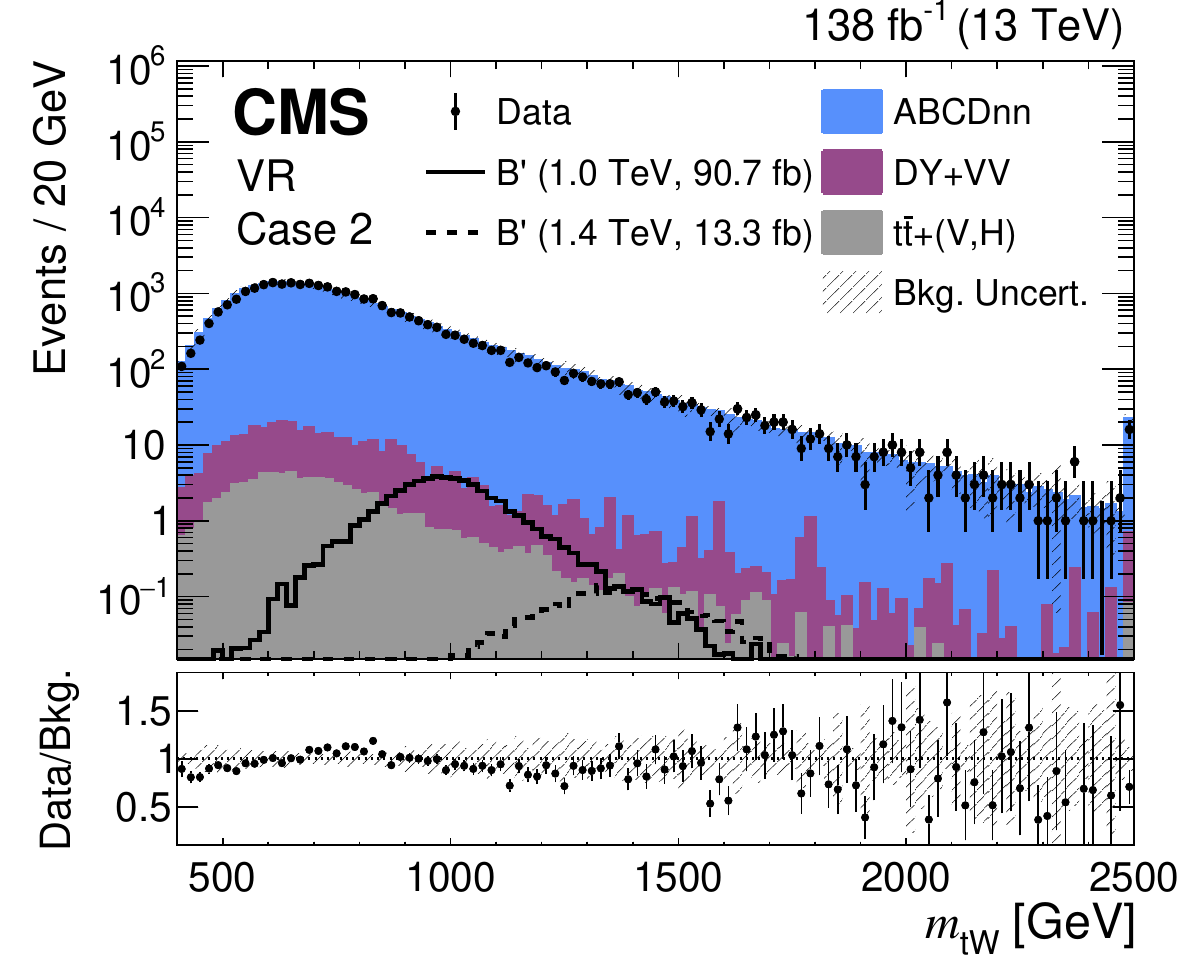}
  \includegraphics[width=0.49\textwidth]{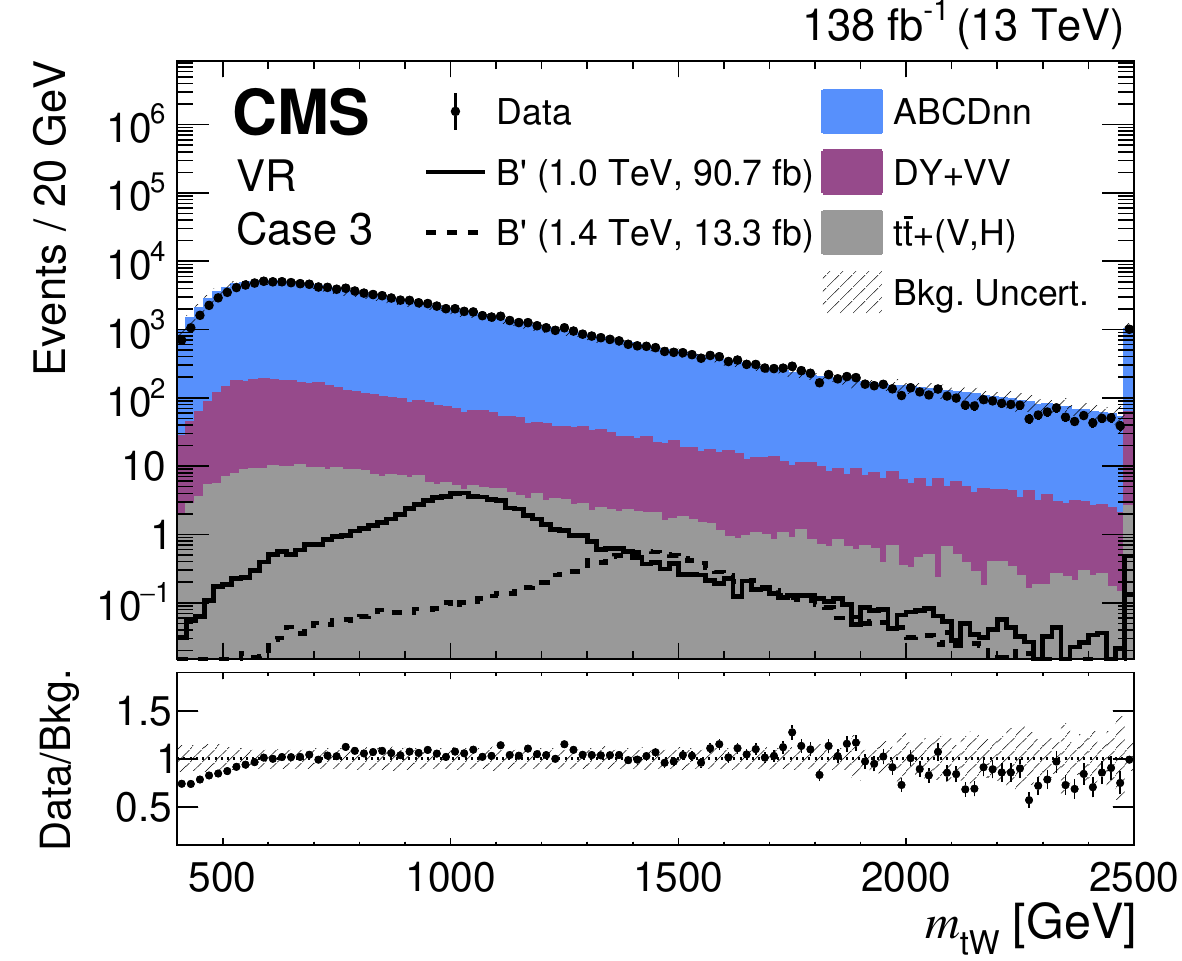}
  \includegraphics[width=0.49\textwidth]{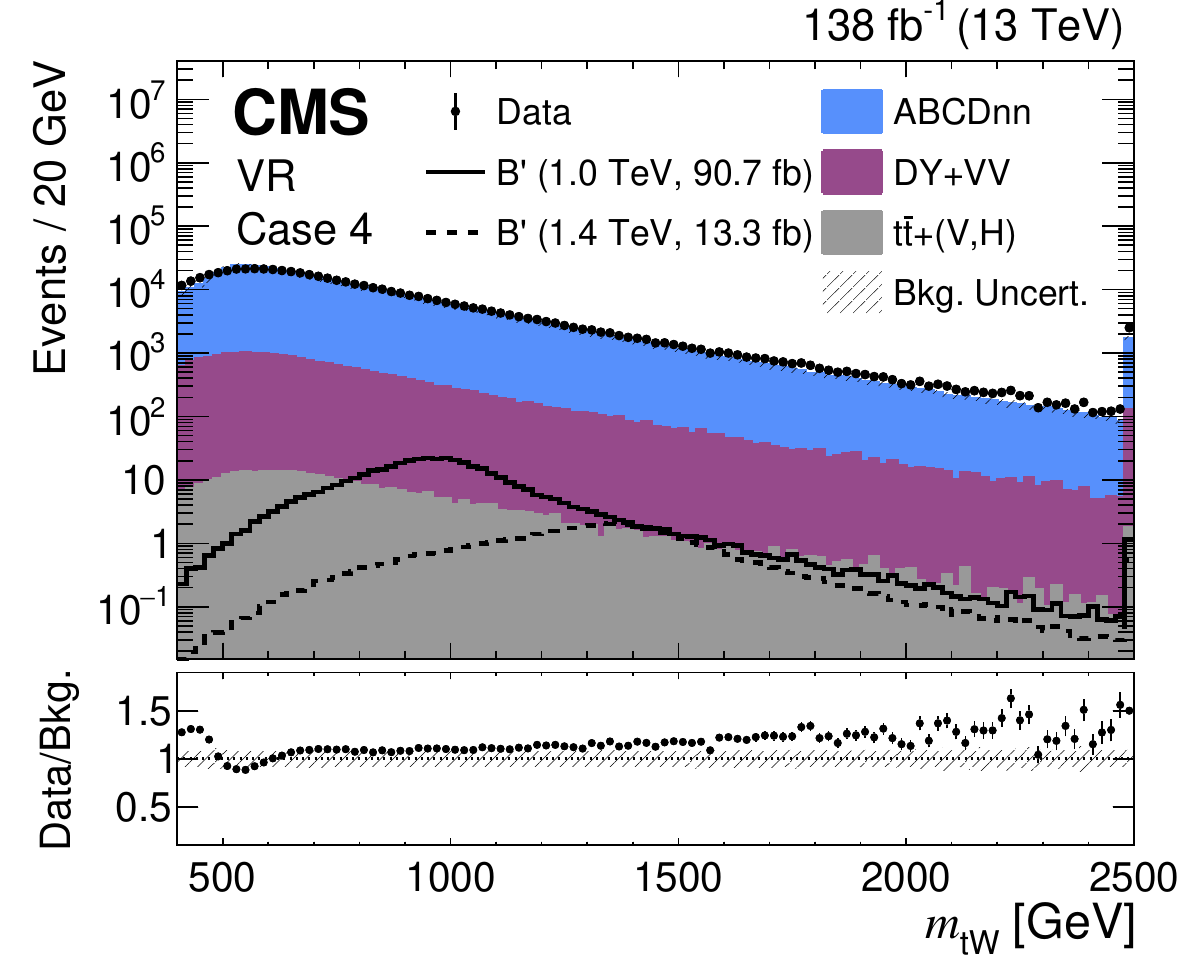}
  \caption{Distributions of \mtw for VR events in Case 1 (upper left) through Case 4 (lower right).
  Observed data are shown as black markers.
  Predicted $\PQb\Pq\PQBpr$ quark signals with masses of 1.0 and 1.4\TeV are shown as solid and dashed lines, respectively, normalized to the predicted cross section for singlet $\PQb\Pq\PQBpr$ production with $\Gamma/m_{\PQBpr}=5\%$.
  Background estimates are displayed as filled histograms, with the ABCDnn prediction shown prior to application of the closure correction.
  The lower panels show the ratio of the data to the background prediction.
  Statistical and systematic uncertainties in the background estimate are indicated by hatched bands.
  }
  \label{fig:allV}
\end{figure*}

Figure~\ref{fig:allV} shows the \mtw distributions in the VR for each reconstruction case.
The background estimate includes simulated minor background processes and the ABCDnn prediction for the major background processes. 
The \ParticleNet tag information from the major background simulation is used to separate the ABCDnn prediction into the four reconstruction cases, as described in Section~\ref{sec:Reconstruction}.
Overall, the agreement between the data and the background prediction demonstrates that the ABCDnn method successfully models the major backgrounds, although residual corrections are required. 
These corrections are derived empirically and applied to improve the prediction.

The bin-by-bin discrepancies between observed data and prediction in the VRs are evaluated in the two-dimensional (2D) distribution of \mtw and \ST to derive a closure correction, where both variables are taken from the ABCDnn prediction.
Minor backgrounds modeled from simulation are subtracted from the data before these discrepancies are computed.
The \ST axis is divided into variable-width bins to ensure that all bins contain nonzero ABCDnn predictions and that a bin boundary is placed at 850\GeV.
For Cases~1 and~2, bins in the parameter space exclusive to the SR with $\ST>850\GeV$ and $\mtw>700\GeV$, where the signal contribution is non-negligible, are assigned values based on the discrepancies observed in the corresponding bins of CR~$B$.
The ABCDnn prediction is corrected by applying the measured discrepancy to each bin.
The corrected 2D distributions are then projected into one-dimensional (1D) distributions of \mtw, and smoothed using the LOWESS technique~\cite{LOWESS1,LOWESS2}.
This smoothing is applied in all regions where the correction is derived from the VR rather than from CR~$B$, to reduce the dependence of the prediction on statistical fluctuations in the data. After this correction, agreement between data and prediction in the four SR distributions of \mtw is consistent with unity within uncertainties. The uncertainty associated with the closure correction procedure is described in Section~\ref{sec:Systematics}. The ABCDnn method provides comparable or better agreement between the distributions of data and prediction, while greatly simplifying the estimation of the uncertainty associated with the dominant backgrounds by considering only five (four) sources for Cases 1 and 2 (Cases 3 and 4), as summarized in Section~\ref{sec:Systematics}.

\section{Systematic uncertainties}\label{sec:Systematics}

Systematic uncertainties affect both the normalization and the shape of the predicted background and signal distributions, and are summarized in Table~\ref{table:Systematics}.
The impact of shape-based uncertainties in processes modeled with simulation is evaluated by varying the corresponding inputs within their uncertainties.
Systematic uncertainties are correlated across the data-taking periods and lepton flavors unless specified otherwise.

\begin{table*}[th]
  \centering
  \topcaption{
  A summary of the systematic uncertainties impacting the search.
  Unless otherwise specified, all uncertainties are implemented as rate and shape variations using alternate template histograms.
  The numerical values represent the prefit rate impact and are given as ranges across the four reconstruction cases for either the 1.4\TeV signal or the ABCDnn background prediction, unless otherwise specified.
  The functional form in the third column indicates the quantities on which the uncertainty magnitude depends, where \SD~denotes one standard deviation.
  The final column indicates the affected predictions: simulated samples (including signal), ``minor'' refers to simulated minor background processes, and ``ABCDnn'' refers to the   major background predicted using the ABCDnn method.
  }
  \label{table:Systematics}
  \begin{scotch}{lccc}
      Uncertainty                     & Value      & Formula                              & Affected Samples \\
      \hline                                             
      Integrated luminosity           & 0.73\%      & rate only                            & Simulation      \\
      Pileup                          & $<$1\%     & $\sigma_{\text{inel}}\pm$4.6\%       & Simulation      \\
      \Pe trigger                   & $<$1\%     & $\pm$\SD(\pt,$\eta$)              & Simulation      \\
      \PGm trigger                  & $<$1\%     & $\pm$\SD(\pt,$\eta$)              & Simulation      \\
      L1 trigger timing          & 1\%        & $\pm$\SD(\pt,$\eta$)              & Simulation      \\
      Lepton scale factors            & $<$1\%     & $\pm$\SD(\pt,$\eta$)              & Simulation      \\
      Jet energy scale                & $\leq$2\% & $\pm$\SD(\pt,$\eta$)              & Simulation      \\
      Jet energy res.                 & $\leq$1\% & $\pm$\SD($\eta$)                  & Simulation      \\
      Pileup jet ident.               & 1\%        & $\pm$\SD(\pt,$\eta$)              & Simulation      \\
      \PQb tag: {\PQb}/{\PQc}         & $\leq$3\% & $\pm$\SD(\pt)                     & Simulation      \\
      \PQb tag: light                 & $\leq$1\% & $\pm$\SD(\pt)                     & Simulation      \\
      Renormalization scale           & 4--11\% (minor) & $\mu_\text{R}\times 0.5,\mu_\text{R}\times 2$   & Simulation      \\
      Factorization scale             & 3--5\% (minor)  & $\mu_\text{F}\times 0.5,\mu_\text{F}\times 2$   & Simulation      \\
      PDF                             & 1--6\% (minor)  & $\pm$RMS (Hessian)         & Simulation      \\
      \ParticleNet: \PQt tag  & 9\% (signal) & $\pm$\SD(\pt)              & all      \\
      \ParticleNet: \PW tag   & 14\% (signal) & $\pm$\SD(\pt)             & all      \\
      Training accuracy               & 1--25\%    & $\pm$ $\%(\text{CR}_\text{pred} - \text{CR}_\text{data})$ & ABCDnn \\
      VR closure                      & 1--3\%     & $\text{corr.}\times0,\text{corr.}\times2$  & ABCDnn \\
      ABCDnn smoothing                & 1--2\%     & $\pm \Delta(\text{2D} - \text{1D})$   & ABCDnn \\
      ABCDnn normalization            & 2--8\%     & rate only                             & ABCDnn \\
  \end{scotch}
\end{table*}

\begin{itemize}

\item
The uncertainty in the integrated luminosity for the combined 2016--2018 data-taking period is 0.73\%~\cite{CMS-PAS-LUM-20-001}.

\item
The uncertainty in the pileup weighting for simulation is evaluated by varying the total inelastic cross section $\sigma_\text{inel}=69.2\unit{mb}$ by $\pm4.6\%$~\cite{CMS:2018mlc}.

\item
Single electron and single muon trigger efficiency corrections are applied to simulation to account for differences between data and simulation.
The corresponding uncertainties are applied independently for each lepton flavor and data-taking period as functions of \pt and $\eta$.

\item
An L1 trigger timing correction is applied in the 2016 and 2017 data-taking periods to account for a gradual shift in the ECAL L1 trigger timing in the region $\abs{\eta}>2.0$.
The uncertainty in this correction is applied as a function of \pt and $\eta$.

\item
Lepton scale factors are applied to correct for differences in the reconstruction, identification, and isolation efficiencies between data and simulation.
The associated uncertainties are applied independently for each lepton flavor and correction component as functions of \pt and $\eta$.

\item
Jet energy scale and resolution uncertainties affect the momenta of both small- and large-radius jets.
These variations are propagated to \ptmiss and all jet-dependent observables.
The resulting varied histograms are smoothed using the LOWESS algorithm.
The uncertainties are treated independently for each data-taking period.

\item
The uncertainty in the pileup jet identification scale factor is applied as a function of \pt and $\eta$.

\item
The \textsc{DeepJet} \PQb-tagging corrections are applied to simulation to match the tagging efficiencies observed in data.
The uncertainties are separated into four components: b/c jet tagging is treated separately from light-quark/gluon misidentification~\cite{CMS-DP-2018-058}; 
and for each of these there is one component correlated across data-taking periods and one that is uncorrelated.

\item
Scale uncertainties are used to account for missing higher-order corrections.
The renormalization and factorization scales ($\mu_\text{R}$ and $\mu_\text{F}$) are varied independently up and down by factors of two for each simulated process group defined in Section~\ref{sec:DataMC}.
The resulting variations are evaluated on a bin-by-bin basis in \mtw.
For the signal, the variations are normalized to preserve the total yield, such that the uncertainty affects only the shape.

\item
The PDF uncertainties are correlated across all samples.
The NNPDF~3.1 Hessian uncertainties are evaluated bin-by-bin in \mtw using 100 eigenvector variations and combined in quadrature.
For the signal, the normalization effect is removed, such that only shape variations are considered.

\item
The \ParticleNet scale factors are applied to correct the efficiencies for tagging large-radius jets originating from generator-level \PQt quarks or \PW bosons.
The uncertainties are applied as functions of the jet \pt.
The \PQt tagging efficiency affects only events in Case~1, and the \PW tagging efficiency affects only events in Case~2.
These uncertainties apply to both simulation and ABCDnn predictions.

\item
The ABCDnn training accuracy uncertainty accounts for differences between data (after subtraction of minor background predictions) and the ABCDnn prediction in the CRs.
It is derived from the average percentage difference between data and the ABCDnn prediction in the CRs $A$, $B$, and $C$, evaluated in each bin of \mtw.
Two alternate templates are formed by varying the prediction up and down by this amount.
This uncertainty is evaluated separately for each reconstruction case as a function of \mtw and \ST.
In Table~\ref{table:Systematics}, this uncertainty is defined as the relative difference $(\text{CR}_\text{pred} - \text{CR}_\text{data})/\text{CR}_\text{pred}$, expressed in percent.

\item
The VR closure uncertainty accounts for differences arising from the propagation of the ABCDnn transfer function, determined during training in the CRs, to the SR.
It is evaluated by forming alternate templates in which the closure correction is either removed or doubled in each bin of \mtw and \ST.
In Table~\ref{table:Systematics}, this uncertainty is denoted as $\text{corr.}~\times0,\text{corr.}\times2$.

\item
The ABCDnn smoothing uncertainty is evaluated by comparing the corrected 1D \mtw distributions to those obtained from smoothing the 2D $(\mtw,\ST)$ distributions with a kernel method and projecting onto \mtw.
Alternate templates are formed by varying the background prediction by this difference.
In Table~\ref{table:Systematics}, this uncertainty is denoted as $\Delta(\text{2D} - \text{1D})$.

\item
The ABCDnn normalization uncertainty accounts for the limited event counts used to compute the normalization factor in Eq.~\ref{eq:norm}, as well as the differences between the predicted and observed event yields in the VRs.
It is applied as an independent normalization uncertainty in each reconstruction case.

\end{itemize}

The uncertainties that have the largest impact on signal sensitivity are those associated with the \ParticleNet tagging scale factors.
The ABCDnn background prediction uncertainty in Cases 1 and 2 also contributes significantly to the overall uncertainty.

\section{Results}\label{sec:Results}

The strength of the \PQBpr quark signal hypothesis is tested using a binned maximum likelihood fit to the \mtw distribution in region $D$, with one distribution per reconstruction case.
Cases 3 and 4 are fitted simultaneously with Cases 1 and 2 to constrain nuisance parameters related to simulated processes.
The histograms are binned in 10\GeV intervals, and bins at high \mtw values are merged to ensure that the total background in each bin has a statistical uncertainty below 20\%.
Signal distributions for \PQBpr quark masses of 0.9 and 1.1\TeV are created by fitting a double-sided Crystal Ball function, defined in Ref.~\cite{Gaiser:1982}, to the simulated signal \mtw distributions in each case.
The function parameters are interpolated between neighboring simulated mass points, and the resulting formulas for the 0.9\TeV and 1.1\TeV \PQBpr quark masses are used to construct the \mtw histograms.
The same procedure is used to model the systematic uncertainty variations.
The CMS statistical analysis tool \COMBINE~\cite{CMS:2024onh}, based on the \textsc{RooFit}~\cite{Verkerke:2003ir} and \textsc{RooStats}~\cite{Moneta:2010pm} frameworks, is used to perform the fit.
Uncertainties due to limited event counts in simulation are included as Poisson-distributed nuisance parameters using the Barlow--Beeston method~\cite{BBLITE1,BBLITE2}.
Systematic uncertainties listed in Table~\ref{table:Systematics} are included as lognormal-distributed nuisance parameters if they affect only the normalization of the background or signal predictions.
Systematic uncertainties that affect the shapes of template distributions are implemented through template morphing with Gaussian-distributed nuisance parameters~\cite{BBLITE2}.
The fit model is validated using fits to data in the VRs and using pseudo-data generated from the background estimate in the SR, with and without injected \PQBpr signal contributions.
No bias in fitting injected signal is observed for either the $\PQb\Pq\PQBpr$ or $\PQt\Pq\PQBpr$ signals over a wide range of signal mass.

\begin{figure*}[htbp]
  \centering
  \includegraphics[width=0.49\textwidth]{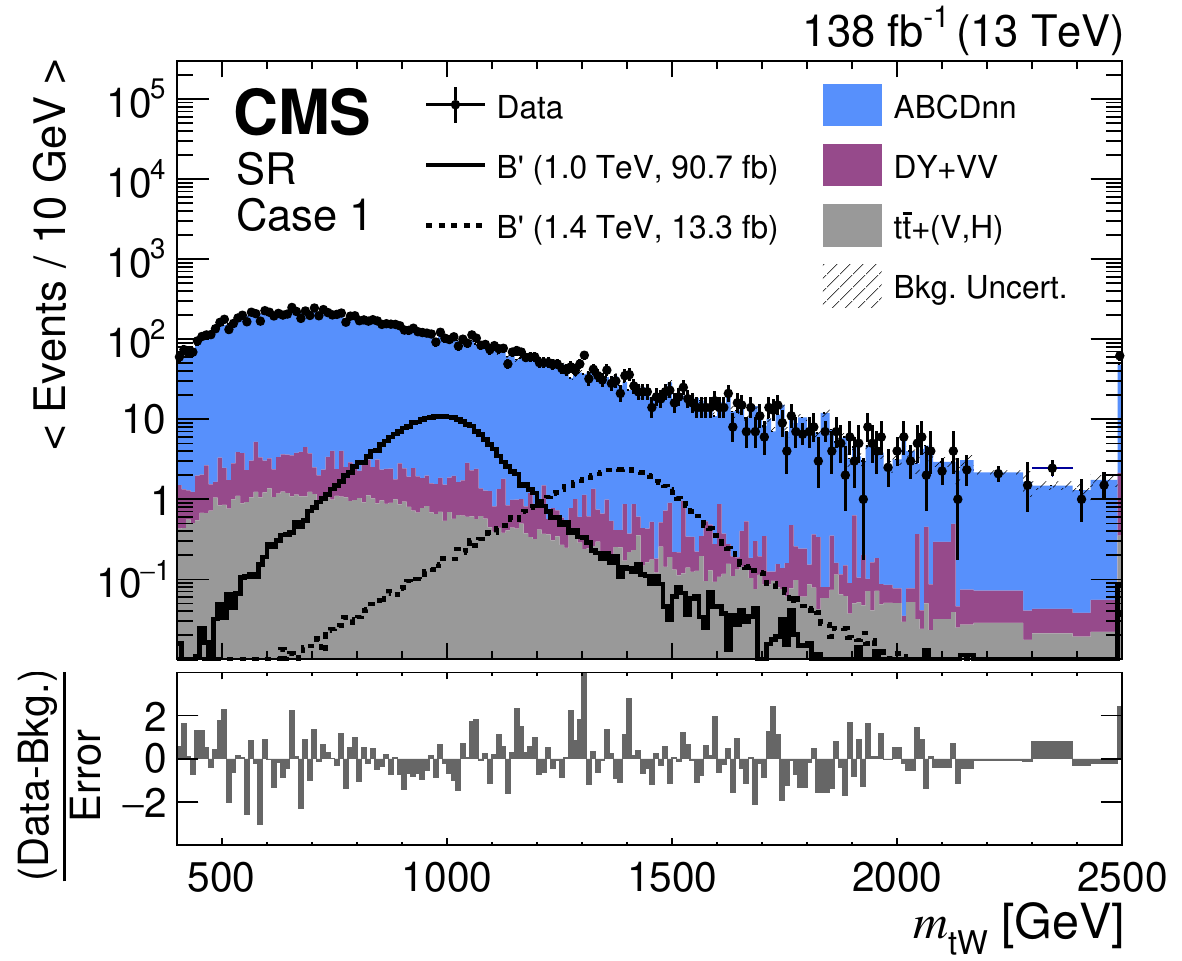}
  \includegraphics[width=0.49\textwidth]{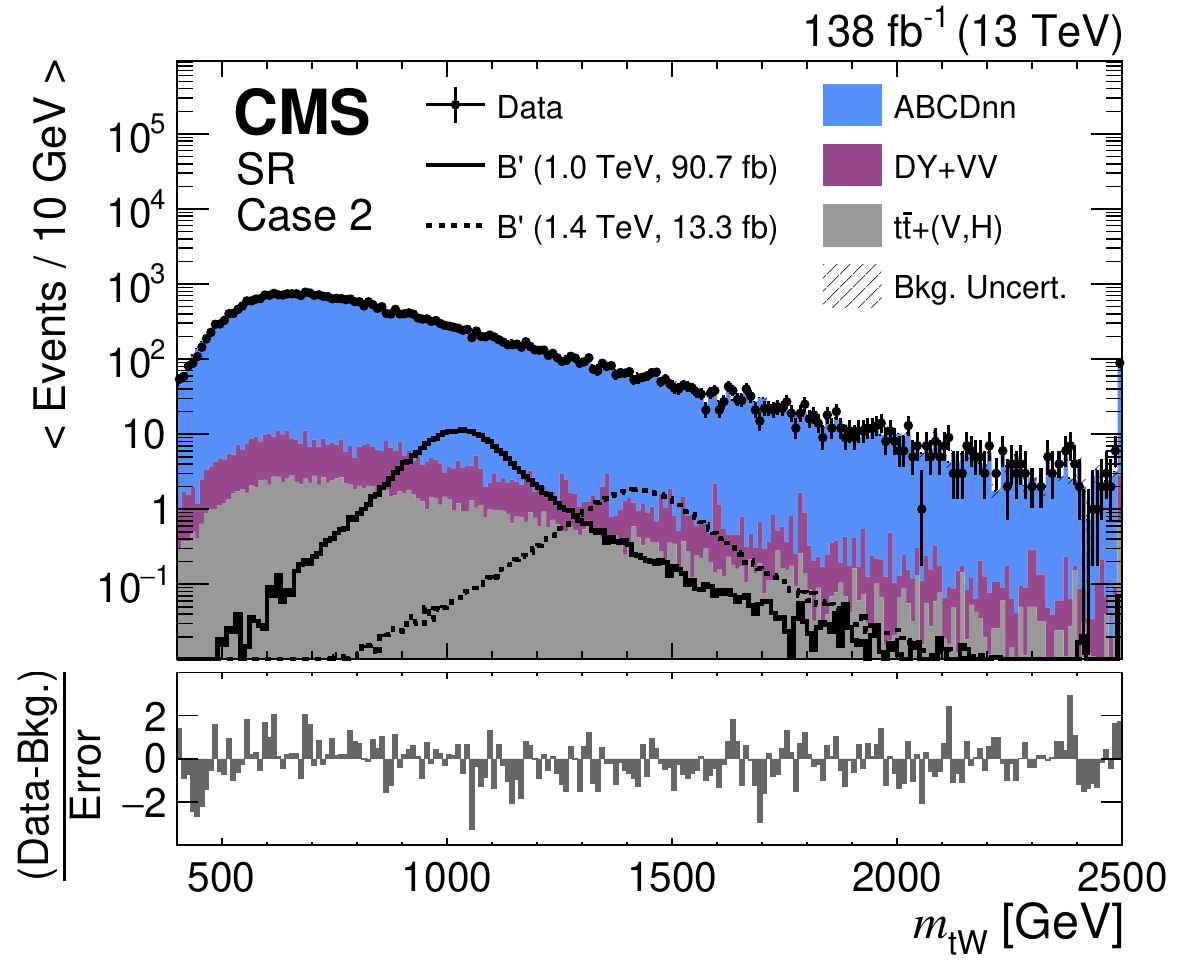}\\
  \includegraphics[width=0.49\textwidth]{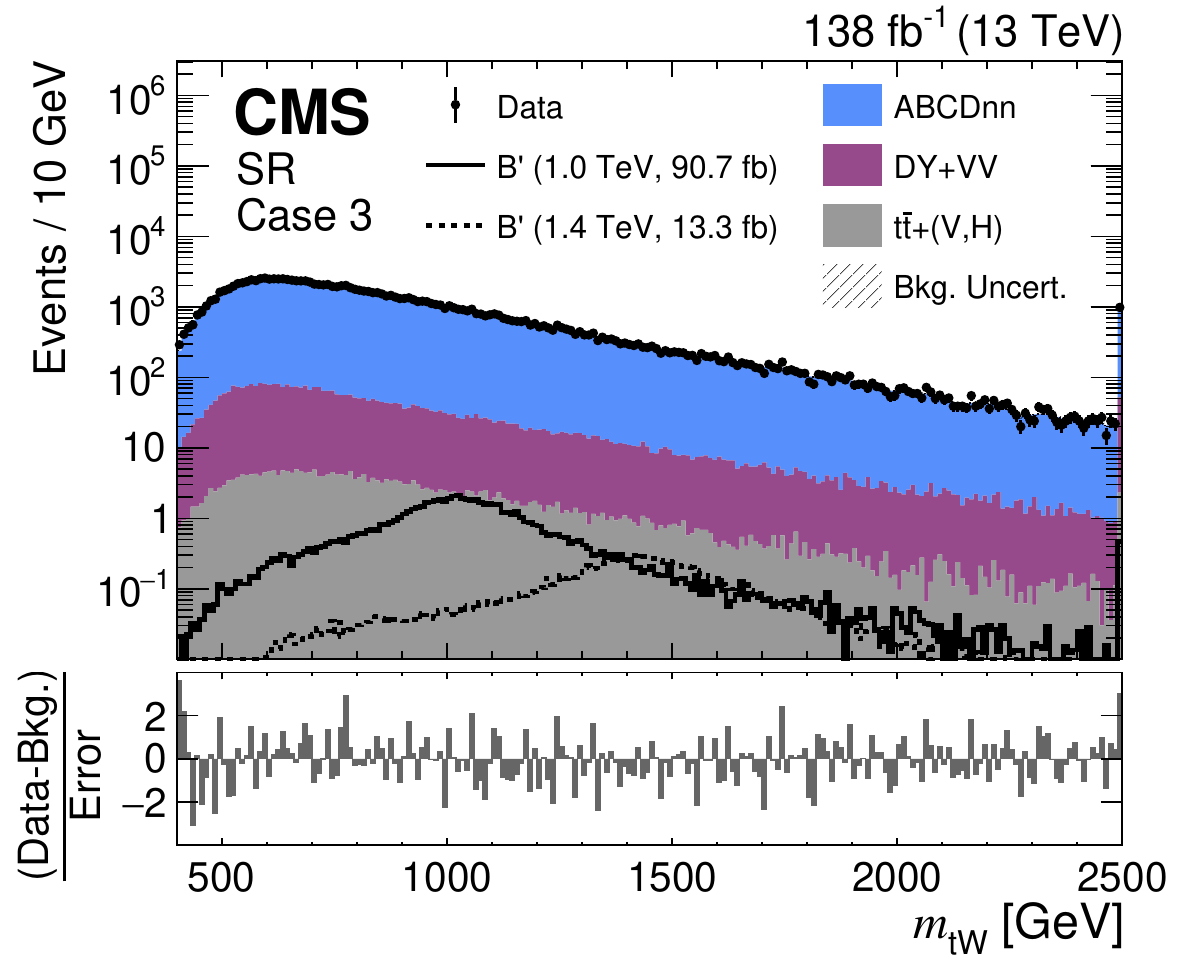}
  \includegraphics[width=0.49\textwidth]{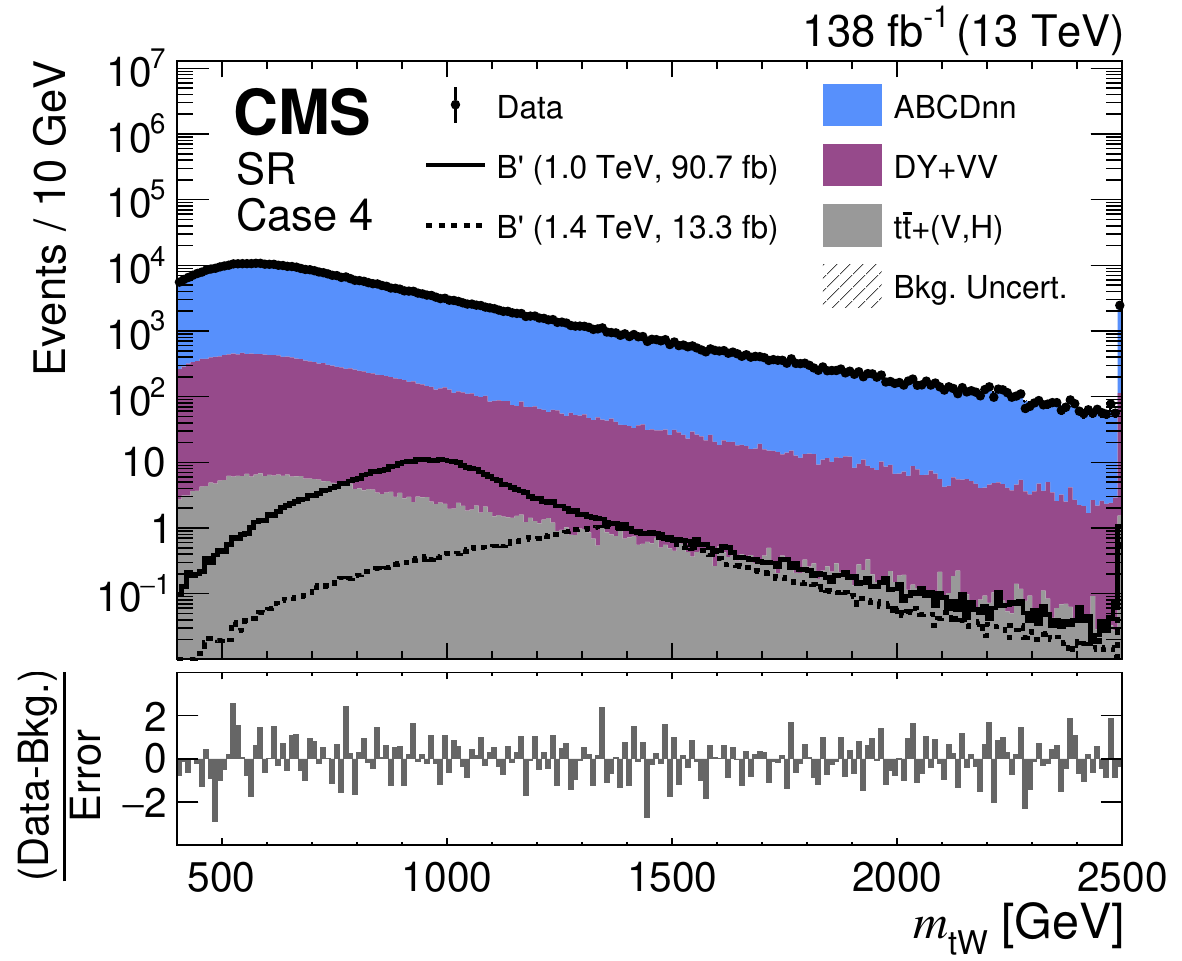}
  \caption{Distributions of \mtw for events in the SR in Case 1 (upper left) through Case 4 (lower right).
  The observed data are shown as black markers.
  Predicted $\PQb\Pq\PQBpr$ quark signals with masses of 1.0 and 1.4\TeV are shown as the solid and dashed lines, respectively, normalized to the predicted cross section for singlet $\PQb\Pq\PQBpr$ production with $\Gamma/m_{\PQBpr}=5\%$.
  The best-fit background prediction from a background-only fit to data is shown as the filled histograms.
  The lower panels show the difference between the data and the background prediction, divided by the total uncertainty in the background estimate.
  }
  \label{fig:allD}
\end{figure*}

Figure~\ref{fig:allD} shows the reconstructed \mtw distributions in the SR, separated by reconstruction case, after the background-only fit to data.
No significant excess of data over the SM background estimate is observed in any reconstruction case, and the background-only fit yields a $p$-value of 0.15 based on a saturated $\chi^2$ test described in Ref.~\cite{CMS:2024onh}.

\begin{figure*}[htbp]
  \centering
  \includegraphics[width=0.49\textwidth]{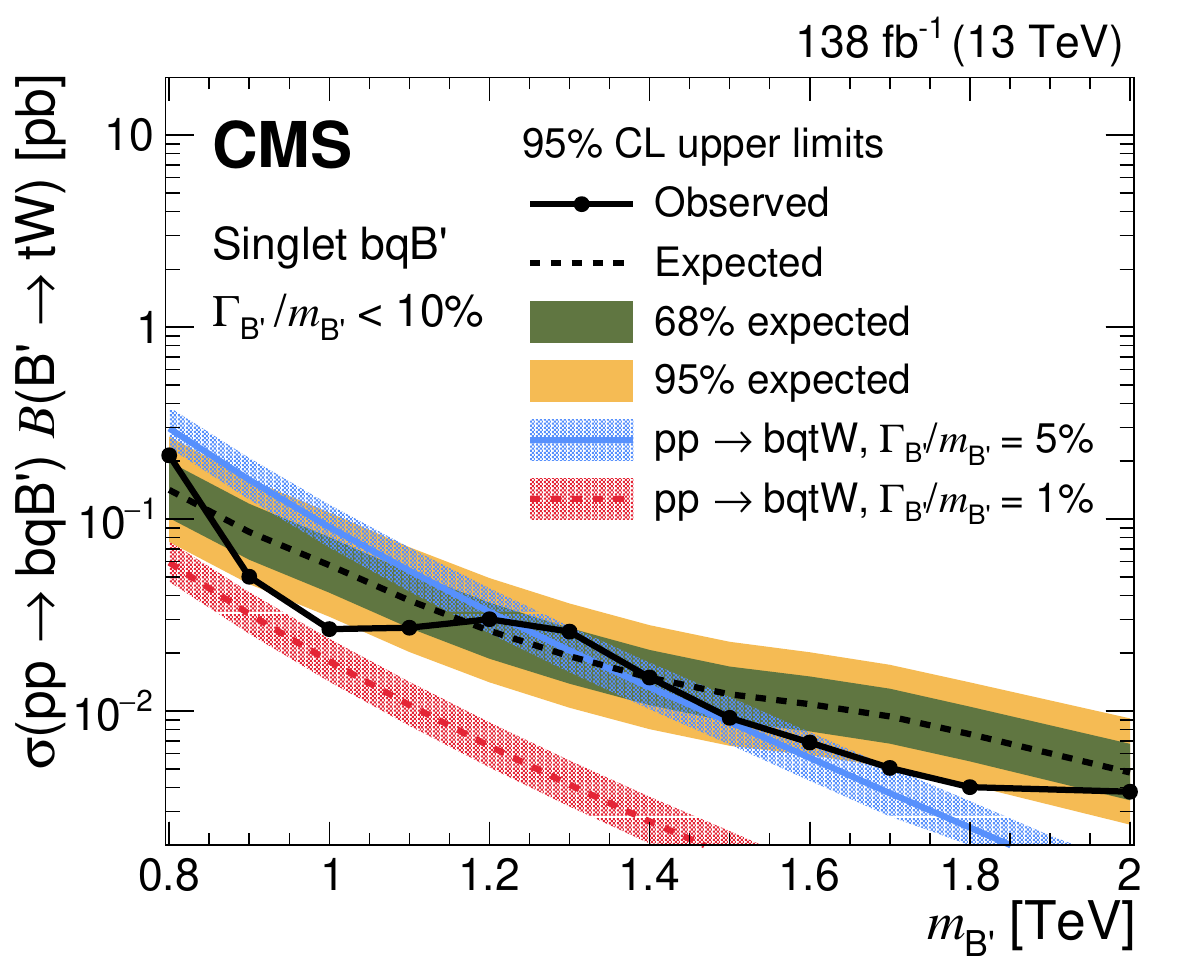}\\
  \includegraphics[width=0.49\textwidth]{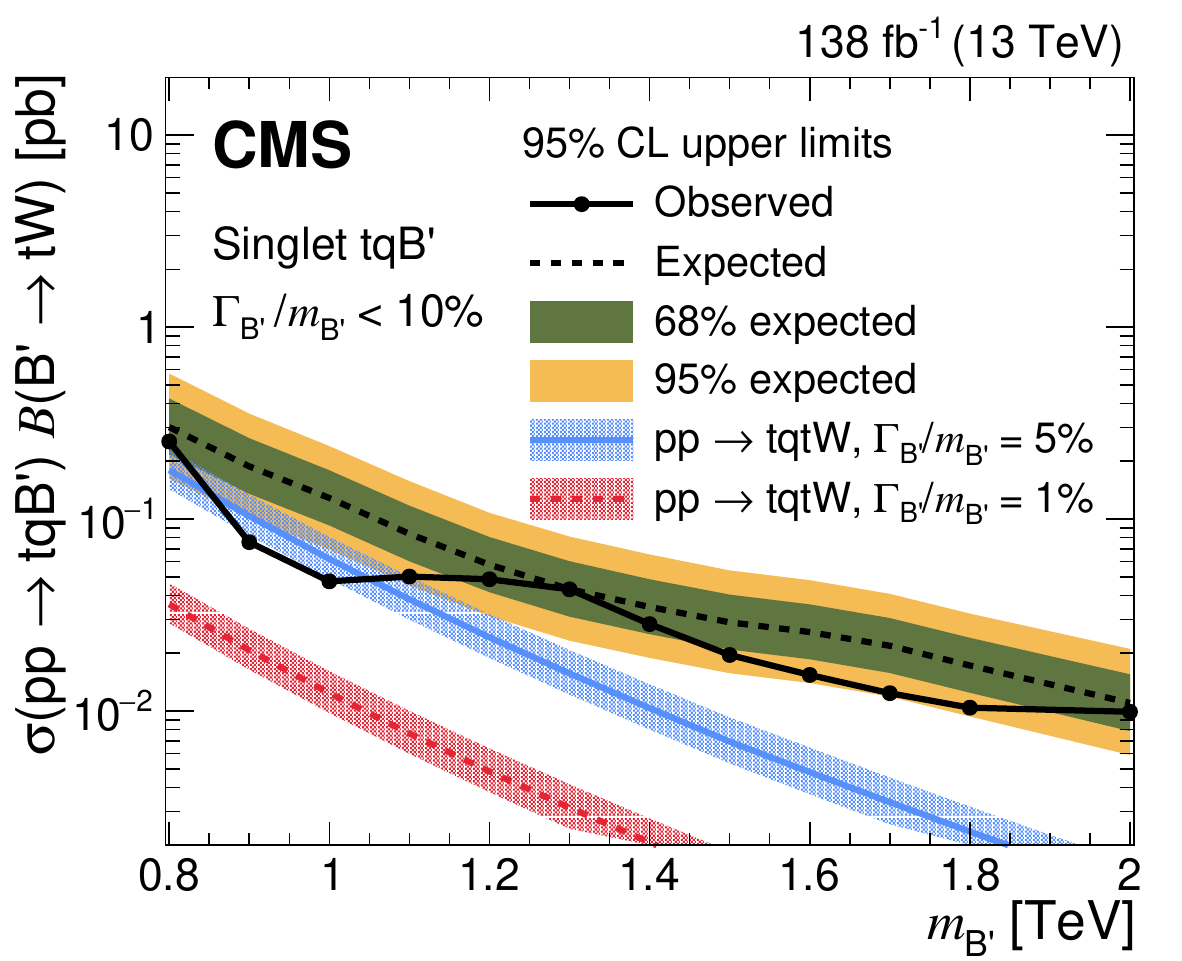}
  \includegraphics[width=0.49\textwidth]{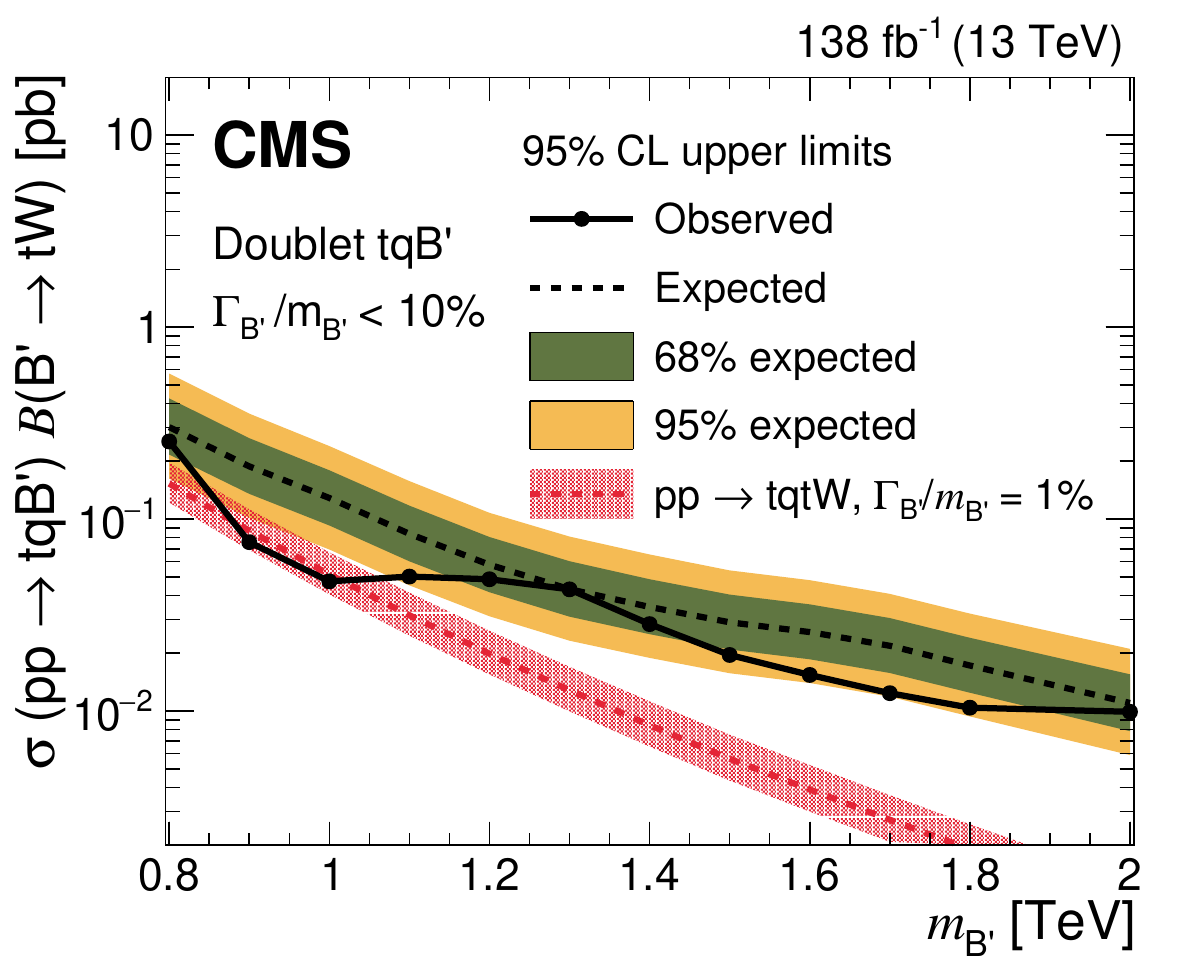}
  \caption{Observed (solid black lines) and expected (dashed lines) 95\% \CL upper limits on the product of the production cross section and branching fraction $\mathcal{B}(\tW)$
  for singlet $\PQb\Pq\PQBpr$ production (upper), singlet $\PQt\Pq\PQBpr$ production (lower left), and up-type doublet $\PQt\Pq\PQBpr$ production (lower right), as functions
  of $m_{\PQBpr}$. Predicted cross sections for various \PQBpr quark relative widths are shown as dashed red and solid blue lines.
  The bands around the predicted cross sections represent the associated energy scale and PDF uncertainties.}
  \label{fig:limits}
\end{figure*}

Upper limits at 95\% confidence level (\CL) on the product of the production cross section and the branching fraction for $\PQb\Pq\PQBpr$ or $\PQt\Pq\PQBpr$ production with $\PQBpr\to\tW$ decay are computed using a profile likelihood test statistic in the asymptotic approximation~\cite{Cowan}, with the \CLs method~\cite{CLS2, CLS1}, and shown in Fig.~\ref{fig:limits}.
For production in association with a \PQb quark, singlet \PQBpr quarks with a 5\% relative decay width are excluded for masses below 1.23\TeV (1.35\TeV expected).
The deficit observed for the $m_{\PQBpr}=1.0\TeV$ hypothesis has a local (global) significance of 2.5 (1.4) s.d. 
This deficit arises from the overpredictions visible in the Case 1 and Case 2 \mtw distributions of Fig.~\ref{fig:allD} in the range of approximately 900--1200\GeV.
The overprediction in this range of \mtw is statistically significant in Case 2. 
For production in association with a \PQt quark, the observed limits range from 230\unit{fb} at $m_{\PQBpr}=0.8\TeV$ to 10\unit{fb} at $m_{\PQBpr}=2.0\TeV$.
The overpredictions noted for Cases 1 and 2 near $\mtw=1.0\TeV$ lead to the exclusion of the $m_{\PQBpr}=0.9$ and 1.0\TeV hypotheses for \PQt-associated production.
These exclusions apply to a singlet \PQBpr quark with a 5\% relative decay width and to a doublet \PQBpr quark with a 1\% relative decay width.
In this production mode, the deficit at 1.0\TeV has a local (global) significance of 3.6 (2.6) \SD because of the wider \mtw resolution for $\PQt\Pq\PQBpr$ signals.
This search provides the most stringent sensitivity to date to production of single \PQBpr quarks with narrow decay widths.
With the addition of data collected in 2017--2018 and improved techniques such as the \ParticleNet jet identification algorithm, this search is able to probe 
smaller cross sections and coupling factor values compared to earlier CMS searches.

\begin{figure*}[htbp]
  \centering  
  \includegraphics[width=0.49\textwidth]{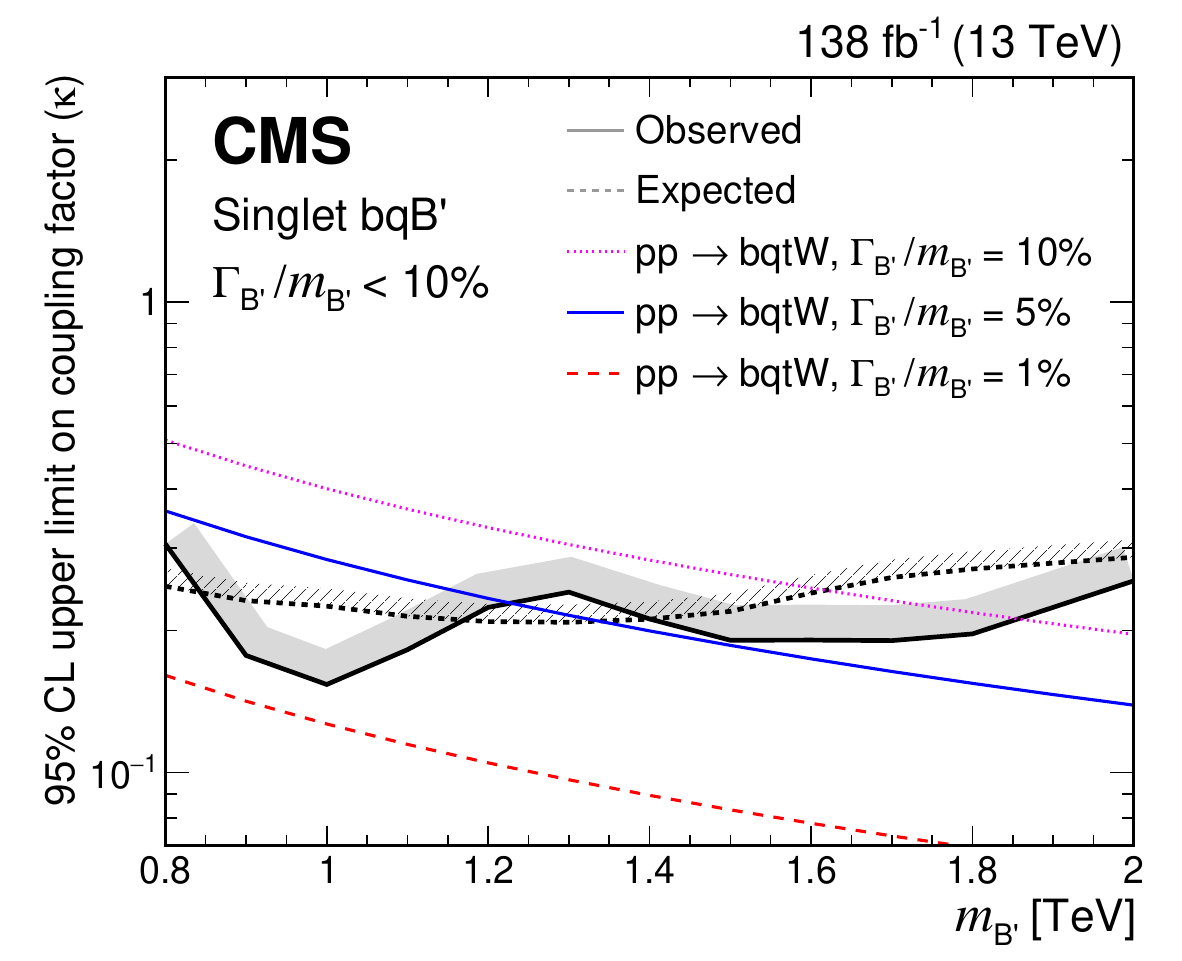}\\
  \includegraphics[width=0.49\textwidth]{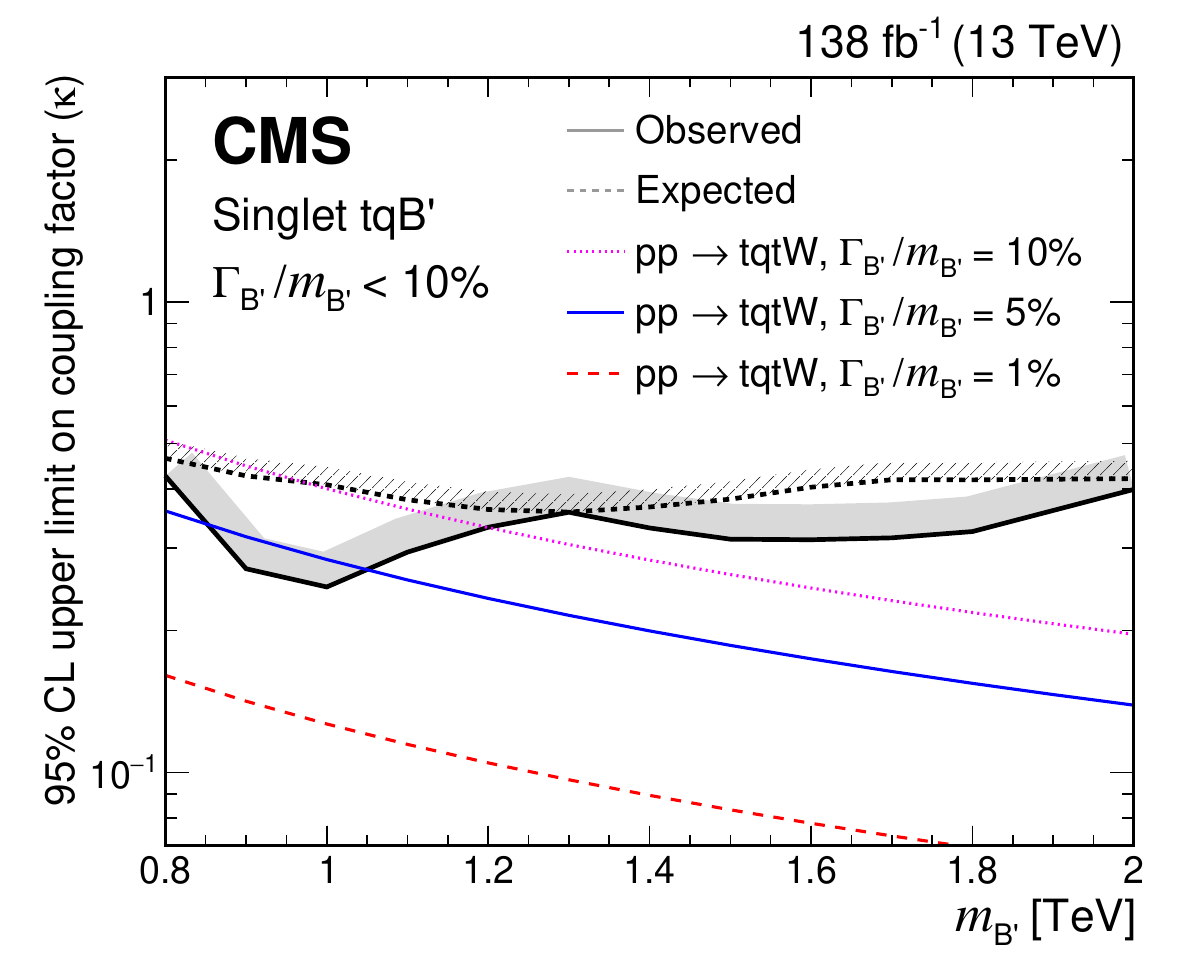}
  \includegraphics[width=0.49\textwidth]{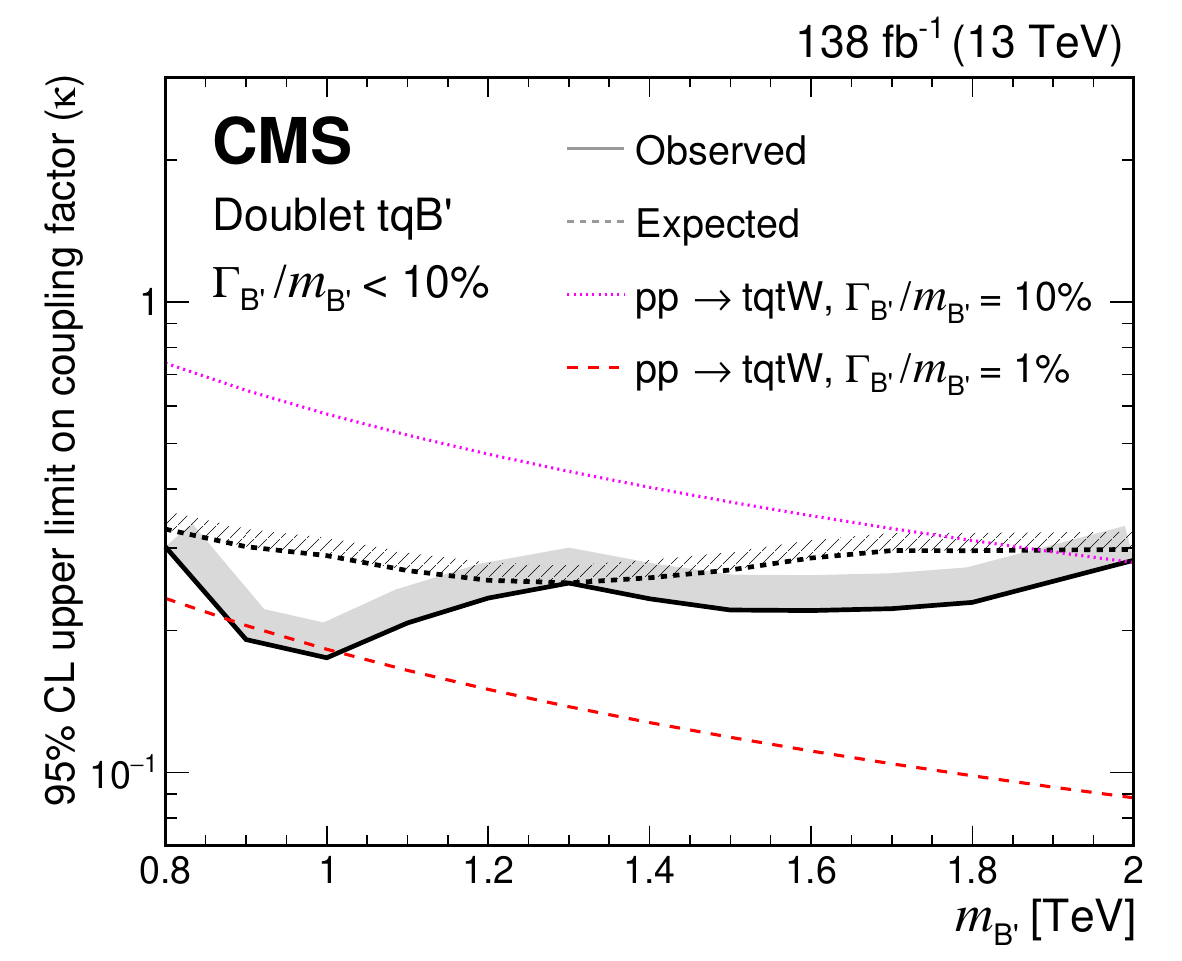}
  \caption{Observed (solid black lines) and expected (dashed lines) 95\% \CL upper limits on the coupling factor $\kappa$ for singlet $\PQb\Pq\PQBpr$ production (upper),
  singlet $\PQt\Pq\PQBpr$ production (lower left), and up-type doublet $\PQt\Pq\PQBpr$ production (lower right), as functions of $m_{\PQBpr}$.
  The hatched and shaded regions indicate that the limits are upper bounds on the coupling factor.
  Predicted coupling values for various \PQBpr quark relative widths are shown as dashed red, solid blue, and dotted magenta lines.}
  \label{fig:coupling}
\end{figure*}

Figure~\ref{fig:coupling} presents limits on the \PQBpr quark coupling factor $\kappa$, defined in Section~\ref{sec:DataMC}.
The predicted couplings as a function of the \PQBpr quark mass can be extended to a 10\% relative decay width.
At this width, the natural width of the \PQBpr quark becomes comparable to the resolution of the \mtw distribution, and the narrow-width signal simulation used in this search may no longer provide an accurate description of the signal.
Therefore, the coupling limits are only interpreted in the region where the observed upper limits of the coupling factor lie below the values corresponding to $\Gamma/m_{\PQBpr}= 10\%$.
For singlet scenario of $\PQb\Pq\PQBpr$ production, the upper limits on the coupling factor range from 0.15 to 0.31.
The observed coupling factor limits become consistent with a \PQBpr quark with a 10\% relative decay width for masses above 1.8\TeV.
For $\PQt\Pq\PQBpr$ production, the upper limits on the coupling factor range from 0.25 to 0.43 in the singlet scenario and from 0.18 to 0.30 in the doublet scenario.
In the singlet case, the observed coupling limits become consistent with a \PQBpr quark with a 10\% relative decay width for masses above 1.2\TeV.

\clearpage
\section{Summary}\label{sec:Summary}
A search has been presented for the production of a single vector-like \PQBpr quark that decays to a \PQt quark and a \PW boson via the electroweak interaction. 
Proton-proton collision data collected with the CMS detector at a center-of-mass energy of 13\TeV are used, corresponding to an integrated luminosity of 138\fbinv.
Events are selected with one electron or muon, missing transverse momentum, and at least one large-radius jet well separated from the charged lepton.
A \PQBpr quark candidate is reconstructed and categorized according to the \ParticleNet identification of the large-radius jet.
Minor background processes are modeled using simulation, while the major background contribution in the signal region is modeled from data in five control regions using a neural autoregressive flow network, referred to as ABCDnn.
For $\PQb\Pq\PQBpr$ production, singlet \PQBpr quarks with a 5\% relative decay width and masses between 0.8 and 1.23\TeV are excluded at 95\% confidence level (\CL).
Limits are also set on the product of the production cross section and the $\PQBpr\to\tW$ branching fraction for $\PQt\Pq\PQBpr$ production, excluding the 0.9 and 1.0\TeV mass hypotheses at 95\% \CL for both the 5\%-width singlet and 1\%-width doublet scenarios.
For singlet $\PQb\Pq\PQBpr$ production, coupling factor values above 0.31 are excluded for \PQBpr quark masses in the range 0.8--1.8\TeV.
For $\PQt\Pq\PQBpr$ production, coupling factor values above 0.43 are excluded at 95\% \CL for singlet \PQBpr quark masses in the range 0.8--1.2\TeV, and values above 0.30 are excluded at 95\% \CL for doublet \PQBpr quark masses in the range 0.8--2.0\TeV.
These limits are the most stringent to date for the production cross section and coupling factors of narrow-width \PQBpr quarks.

\begin{acknowledgments}
We congratulate our colleagues in the CERN accelerator departments for the excellent performance of the LHC and thank the technical and administrative staffs at CERN and at other CMS institutes for their contributions to the success of the CMS effort. In addition, we gratefully acknowledge the computing centers and personnel of the Worldwide LHC Computing Grid and other centers for delivering so effectively the computing infrastructure essential to our analyses. Finally, we acknowledge the enduring support for the construction and operation of the LHC, the CMS detector, and the supporting computing infrastructure provided by the following funding agencies: SC (Armenia), BMBWF and FWF (Austria); FNRS and FWO (Belgium); CNPq, CAPES, FAPERJ, FAPERGS, and FAPESP (Brazil); MES and BNSF (Bulgaria); CERN; CAS, MoST, and NSFC (China); MINCIENCIAS (Colombia); MSES and CSF (Croatia); RIF (Cyprus); SENESCYT (Ecuador); ERC PRG and PSG, TARISTU24-TK10 and MoER TK202 (Estonia); Academy of Finland, MEC, and HIP (Finland); CEA and CNRS/IN2P3 (France); SRNSF (Georgia); BMFTR, DFG, and HGF (Germany); GSRI (Greece); MATE and NKFIH (Hungary); DAE and DST (India); IPM (Iran); SFI (Ireland); INFN (Italy); MSIT and NRF (Republic of Korea); MES (Latvia); LMTLT (Lithuania); MOE and UM (Malaysia); BUAP, CINVESTAV, CONACYT, LNS, SEP, and UASLP-FAI (Mexico); MOS (Montenegro); MBIE (New Zealand); PAEC (Pakistan); MSHE, NSC, and NAWA (Poland); FCT (Portugal); MESTD (Serbia); MICIU/AEI and PCTI (Spain); MOSTR (Sri Lanka); Swiss Funding Agencies (Switzerland); MST (Taipei); MHESI (Thailand); TUBITAK and TENMAK (T\"{u}rkiye); NASU (Ukraine); STFC (United Kingdom); DOE and NSF (USA).

\begin{sloppy}
\setlength\emergencystretch{\hsize}
\hyphenation{Rachada-pisek} Individuals have received support from the Marie-Curie program and the European Research Council and Horizon 2020 Grant, contract Nos.\ 675440, 724704, 752730, 758316, 765710, 824093, 101115353, 101002207, 101001205, and COST Action CA16108 (European Union); the Leventis Foundation; the Alfred P.\ Sloan Foundation; the Alexander von Humboldt Foundation; the Science Committee, project no. 22rl-037 (Armenia); the Fonds pour la Formation \`a la Recherche dans l'Industrie et dans l'Agriculture (FRIA) and Fonds voor Wetenschappelijk Onderzoek contract No. 1228724N (Belgium); the Beijing Municipal Science \& Technology Commission, No. Z191100007219010, the Fundamental Research Funds for the Central Universities, the Ministry of Science and Technology of China under Grant No. 2023YFA1605804, the Natural Science Foundation of China under Grant No. 12535004, and USTC Research Funds of the Double First-Class Initiative No.\ YD2030002017 (China); the Ministry of Education, Youth and Sports (MEYS) of the Czech Republic; the Shota Rustaveli National Science Foundation (Georgia); the Deutsche Forschungsgemeinschaft (DFG), among others, under Germany's Excellence Strategy -- EXC 2121 ``Quantum Universe" -- 390833306, and under project number 400140256 - GRK2497; the Hellenic Foundation for Research and Innovation (HFRI), Project Number 2288 (Greece); the Hungarian Academy of Sciences, the New National Excellence Program - \'UNKP, the NKFIH research grants K 131991, K 138136, K 143460, K 143477, K 147557, K 146913, K 146914, K 147048, TKP2021-NKTA-64, and 2025-1.1.5-NEMZ\_KI-2025-00004, and MATE KKP and KKPCs Research Excellence and Flagship Research Groups grants (Hungary); the Council of Science and Industrial Research, India; ICSC -- National Research Center for High Performance Computing, Big Data and Quantum Computing, FAIR -- Future Artificial Intelligence Research, and CUP I53D23001070006 (Mission 4 Component 1), funded by the NextGenerationEU program, the Italian Ministry of University and Research (MUR) under Bando PRIN 2022 -- CUP I53C24002390006, PRIN PRIMULA 2022RBYK7T (Italy); the Latvian Council of Science; the Ministry of Science and Higher Education, project no. 2022/WK/14, and the National Science Center, contracts Opus 2021/41/B/ST2/01369, 2021/43/B/ST2/01552, 2023/49/B/ST2/03273, and the NAWA contract BPN/PPO/2021/1/00011 (Poland); the Funda\c{c}\~ao para a Ci\^encia e a Tecnologia (Portugal); the National Priorities Research Program by Qatar National Research Fund; MICIU/AEI/10.13039/501100011033, ERDF/EU, ``European Union NextGenerationEU/PRTR", projects PID2022-142604OB-C21, PID2022-139519OB-C21, PID2023-147706NB-I00, PID2023-148896NB-I00, PID2023-146983NB-I00, PID2023-147115NB-I00, PID2023-148418NB-C41, PID2023-148418NB-C42, PID2023-148418NB-C43, PID2023-148418NB-C44, PID2024-158190NB-C22, RYC2021-033305-I, RYC2024-048719-I, CNS2023-144781, CNS2024-154769 and Plan de Ciencia, Tecnolog{\'i}a e Innovaci{\'o}n de Asturias, Spain; the Chulalongkorn Academic into Its 2nd Century Project Advancement Project, the National Science, Research and Innovation Fund program IND\_FF\_68\_369\_2300\_097, and the Program Management Unit for Human Resources \& Institutional Development, Research and Innovation, grant B39G680009 (Thailand); the Eric \& Wendy Schmidt Fund for Strategic Innovation through the CERN Next Generation Triggers project under grant agreement number SIF-2023-004; the Kavli Foundation; the Nvidia Corporation; the SuperMicro Corporation; the Welch Foundation, contract C-1845; and the Weston Havens Foundation (USA).
\end{sloppy}
\end{acknowledgments}\section*{Data availability} Release and preservation of data used by the CMS Collaboration as the basis for publications is guided by the  \href{https://doi.org/10.7483/OPENDATA.CMS.1BNU.8V1W}{CMS data preservation, re-use and open access policy}.

\bibliography{auto_generated}
\cleardoublepage \appendix\section{The CMS Collaboration \label{app:collab}}\begin{sloppypar}\hyphenpenalty=5000\widowpenalty=500\clubpenalty=5000\cmsinstitute{University of Tirana and Polytechnic University of Tirana, TIRANA, Albania}
{\tolerance=6000
K.~Tauqeer
\par}
\cmsinstitute{Yerevan Physics Institute, Yerevan, Armenia}
{\tolerance=6000
A.~Gevorgyan\cmsorcid{0000-0003-2751-9489}, A.~Hayrapetyan, V.~Makarenko\cmsorcid{0000-0002-8406-8605}, A.~Tumasyan\cmsAuthorMark{1}\cmsorcid{0009-0000-0684-6742}
\par}
\cmsinstitute{Marietta Blau Institute for Particle Physics, Vienna, Austria}
{\tolerance=6000
W.~Adam\cmsorcid{0000-0001-9099-4341}, L.~Benato\cmsorcid{0000-0001-5135-7489}, T.~Bergauer\cmsorcid{0000-0002-5786-0293}, M.~Dragicevic\cmsorcid{0000-0003-1967-6783}, S.~Gundacker\cmsorcid{0000-0003-2087-3266}, A.K.~Guven\cmsorcid{0009-0004-5670-5138}, P.S.~Hussain\cmsorcid{0000-0002-4825-5278}, M.~Jeitler\cmsAuthorMark{2}\cmsorcid{0000-0002-5141-9560}, N.~Krammer\cmsorcid{0000-0002-0548-0985}, A.~Li\cmsorcid{0000-0002-4547-116X}, D.~Liko\cmsorcid{0000-0002-3380-473X}, M.~Matthewman, J.~Schieck\cmsAuthorMark{2}\cmsorcid{0000-0002-1058-8093}, R.~Sch\"{o}fbeck\cmsAuthorMark{2}\cmsorcid{0000-0002-2332-8784}, M.~Shooshtari\cmsorcid{0009-0004-8882-4887}, M.~Sonawane\cmsorcid{0000-0003-0510-7010}, N.~Van~Den~Bossche\cmsorcid{0000-0003-2973-4991}, W.~Waltenberger\cmsorcid{0000-0002-6215-7228}, C.E.~Wulz\cmsAuthorMark{2}\cmsorcid{0000-0001-9226-5812}
% ...
\par}
$^{1}$Also at Yerevan State University, Yerevan, Armenia\\
$^{2}$Also at Technische Universit\"{a}t Wien, Vienna, Austria\\
% ...
\end{sloppypar}
\end{document}